\documentclass[preprint,prd,aps,showpacs,showkeys,nofootinbib]{revtex4}
\usepackage{amssymb}
\usepackage{amsmath}
\usepackage{multirow}
\usepackage{graphicx}
\usepackage{float}
\usepackage{subfigure} 

\begin{document}

\preprint{}

\title{\boldmath $\bar{B}\to X_s \gamma$ in BLMSSM$^*$}
\footnote{Supported by National Natural Science Foundation of China (11805140, 11347185 and 11905002),
the Scientific and Technological Innovation Programs of
Higher Education Institutions in Shanxi(2017113),
Natural Science Foundation of Shanxi Province(201801D221021 and 201801D221031),
Natural Science Foundation of Hebei Province(A2020201002),
the China Scholarship Council(201906935031)}

\author{
Jian-bin Chen$^1$\footnote[1]{chenjianbin@tyut.edu.cn},
Meng Zhang$^1$, 
Li-Li Xing$^{1}$\footnote[2]{xinglili@tyut.edu.cn},
Tai-Fu Feng$^2$\footnote[3]{fengtf@hbu.edu.cn},
Shu-Min Zhao$^2$\footnote[4]{zhaosm@hbu.edu.cn},
Ke-Sheng Sun$^3$\footnote[5]{sunkesheng@bdu.edu.cn}
}

\affiliation{
$^1$College of Physics and Optoelectronic Engineering, Taiyuan University of Technology, Taiyuan 030024, China\\
$^2$Department of Physics, Hebei University, Baoding, 071002, China\\
$^3$Department of Physics, Baoding University,Baoding, 071000, China}

\begin{abstract}
Applying the effective Lagrangian method,
we study the Flavor Changing Neutral Current $b\to s\gamma$
within the minimal supersymmetric extension of the standard model where baryon
and lepton numbers are local gauge symmetries.
Constraints on the parameters are investigated numerically
with the experimental data on branching ratio of $\bar{B}\to X_s\gamma$.
Additionally, we present the corrections to direct CP-violation in $\bar{B}\rightarrow X_s\gamma$
and time-dependent CP-asymmetry in $B\rightarrow K^*\gamma$.
With appropriate assumptions on parameters, we find the direct CP-violation $A_{CP}$
is very small,
while one-loop contributions to $S_{K^*\gamma}$ can be significant.
\end{abstract}



\maketitle

\section{Introduction}
\label{sec:intro}
Since the Flavor Changing Neutral Current process(FCNC) $b\to s\gamma$
originates only from loop diagrams,
it is very sensitive to new physics beyond the Standard Model(SM).
The updated average data of inclusive $\bar{B}\rightarrow X_s\gamma$ is \cite{PDG}
\begin{eqnarray}
BR(\bar{B}\rightarrow X_s\gamma)_{exp}=(3.40\pm0.21)\times10^{-4}.
\label{exp-bsg}
\end{eqnarray}
and the prediction of SM at next-next-to-leading order (NNLO) is\cite{BrSM1,BrSM2,BrSM3,BrSM4,BrSM5,BrSM6,BrSM7,BrSM8}
\begin{eqnarray}
BR(\bar{B}\rightarrow X_s\gamma)_{SM}=(3.36\pm0.23)\times10^{-4}.
\label{SM-bsg}
\end{eqnarray}
Though the deviation of SM prediction
from experimental results has been almost eliminated in the past few years,
it is helpful to constrain parameters of new physics.

The discovery of Higgs boson on Large Hadron Collider(LHC)
makes SM the most successful theory in particle physics.
Because of the hierarchy problem and missing of gravitational interaction,
it is believed that SM is just an effective approximation of a more fundamental theory
at higher scale.
Among various extensions of SM, supersymmetric models have been studied for decades.

As the simplest extension, the Minimal Supersymmetric Standard Model(MSSM)\cite{MSSM}
solves the hierarchy problem as well as the instability of Higgs boson by introducing a superpartner
for each SM particle.
The Lightest Supersymmetric Particle (LSP) within this frmework also
provides candidates of dark matter as Weakly Interacting Massive Particles (WIMPs).
However the MSSM can not naturely generates tiny neutrino mass which is
needed to explain the observation of neutrino oscillation.
To acquire neutrino masses, heavy majorana neutrinos are introduced in the seesaw mechanism,
which implies that the lepton numbers are broken.
Besides, the baryon numbers are also expected to be broken because of the asymmetry of matter-antimatter in the universe.
The authors of \cite{BLMSSM1,BLMSSM2} present the so called BLMSSM model in which
the baryon and lepton number are local gauged and spontaneously broken at TeV scale.
The experimental bounds on proton decay lifetime is the main motivation of great desert hypothesis.
In BLMSSM, the proton decay can be avoid with discrete
symmetry called matter parity and R-parity\cite{PR597-2015-1-30}

To describe the symmetries of baryon and lepton numbers, gauge group is enlarged to
$SU(3)_C \otimes SU(2)_L \otimes U(1)_Y \otimes U(1)_B \otimes U(1)_L$.
Then corrections to various observations can be induced from new gauge boson and exotic fields
within this scenario.
In ref. \cite{JHEP1411-119}, corrections to anomalous magnetic moment from one loop diagrams
and two-loop Barr-Zee type diagrams are investigated with effective Lagrangian method.
One-loop contributions to $c(t)$ electric dipole moment in CP-Violating BLMSSM is presented in ref. \cite{EPJC-zhao-2017}.
To account for the experimental data on Higgs,
the authors of \cite{NPB-871-2013-223} study the signals of $h\to\gamma\gamma$ and $h\to VV^*(V=Z,W)$
with a 125 GeV Higgs.
In this work, we use the branching ratio to constrain the parameters.
Furthermore, we present the corrections to CP-Violation of $b\to s\gamma$
due to new parameters introduced in this model.

Our presentation is organized as follows. In section \ref{sec-BLMSSM},
we briefly introduce the construction of BLMSSM and the interactions we need for our caculation.
After that, we present the one-loop corrections to branching ratio and CP-Violation
with effective Lagrangian method in section \ref{one-loop-corrections}.
Numerical results are discussed in section \ref{sec-numericial}
and the conclusions is given in section \ref{sec-Conclusions}.

\section{Introduction to BLMSSM}\label{sec-BLMSSM}

The BLMSSM is based on gauge symmetry
$SU(3)_C \otimes SU(2)_L \otimes U(1)_Y \otimes U(1)_B \otimes U(1)_L$.
In order to cancel the anomalies of Baryon number(B),
exotic quarks
$\hat{Q}_4	\sim (3 , 2 , 1/6 , B_4 , 0),
 \hat{U}^c_4 \sim (\bar{3} , 1 , -2/3, - B_4, 0 ),
 \hat{D}^c_4 \sim (\bar{3} , 1 , 1/3 , -B_4 , 0 ),
 \hat{Q}_5^c \sim (\bar{3} , 2 , -1/6, -(1+B_4), 0 ),
 \hat{U}_5   \sim (3, 1 , 2/3 ,  1 + B_4, 0 ),
 \hat{D}_5   \sim (3, 1 , -1/3 , 1 + B_4, 0) $
are introduced.
Baryon number are broken spontaneously after
Higgs superfields
$\hat{\Phi}_B	\sim (  1 , 1 , 0 , 1 , 0 ),
\hat{\varphi}_B	\sim (  1 , 1 , 0 , -1 , 0 )$
acquire nonzero vacuum expectation values(VEVs).
To deal with the anomalies of Lepton number(L), exotic leptons
$\hat{L}_4 	\sim ( 1 , 2 , -1/2 , 0 , L_4),
\hat{E}^c_4	\sim (  1 , 1 , 1 , 0 , -L_4),
\hat{N}^c_4	\sim (  1 , 1 , 0 , 0 , -L_4),
\hat{L}_5^c	\sim (  1 , 2 , 1/2 , 0 , -(3 + L_4)),
\hat{E}_5	\sim (  1 , 1 , -1 , 0 , 3 + L_4),
\hat{N}_5	\sim (  1 , 1 , 0 , 0 , 3 + L_4)$
are introduced, and
$\hat{\Phi}_L	\sim (  1 , 1 , 0 , 0 , -2 ),
\hat{\varphi}_L	\sim (  1 , 1 , 0 , 0 , 2)$
are responsable for the breaking of lepton number\cite{BLMSSM2}.
The superfields
$\hat{X}	\sim ( 1 , 1 , 0 ,2/3 + B_4, 0 ),
\hat{X'}	\sim ( 1 , 1 , 0 ,-(2/3 + B_4), 0)$
which mediate the decay of exotic quarks
are added in this model to avoid their stability.

Given the superfields above, one can construct the superpotential as
\begin{eqnarray}
{\cal W}_{_{BLMSSM}}={\cal W}_{_{MSSM}}+{\cal W}_{_B}+{\cal W}_{_L}+{\cal W}_{_X},
\end{eqnarray}
where ${\cal W}_{_{MSSM}}$ indicates the superpotential of MSSM, and
\begin{eqnarray}
&&{\cal W}_{_B}=\lambda_{_Q}\hat{Q}_{_4}\hat{Q}_{_5}^c\hat{\Phi}_{_B}+\lambda_{_U}\hat{U}_{_4}^c\hat{U}_{_5}
\hat{\varphi}_{_B}+\lambda_{_D}\hat{D}_{_4}^c\hat{D}_{_5}\hat{\varphi}_{_B}+\mu_{_B}\hat{\Phi}_{_B}\hat{\varphi}_{_B}
\nonumber\\
&&\hspace{1.2cm}
+Y_{_{u_4}}\hat{Q}_{_4}\hat{H}_{_u}\hat{U}_{_4}^c+Y_{_{d_4}}\hat{Q}_{_4}\hat{H}_{_d}\hat{D}_{_4}^c
+Y_{_{u_5}}\hat{Q}_{_5}^c\hat{H}_{_d}\hat{U}_{_5}+Y_{_{d_5}}\hat{Q}_{_5}^c\hat{H}_{_u}\hat{D}_{_5}\;,
\nonumber\\
&&{\cal W}_{_L}=Y_{_{e_4}}\hat{L}_{_4}\hat{H}_{_d}\hat{E}_{_4}^c+Y_{_{\nu_4}}\hat{L}_{_4}\hat{H}_{_u}\hat{N}_{_4}^c
+Y_{_{e_5}}\hat{L}_{_5}^c\hat{H}_{_u}\hat{E}_{_5}+Y_{_{\nu_5}}\hat{L}_{_5}^c\hat{H}_{_d}\hat{N}_{_5}
\nonumber\\
&&\hspace{1.2cm}
+Y_{_\nu}\hat{L}\hat{H}_{_u}\hat{N}^c+\lambda_{_{N^c}}\hat{N}^c\hat{N}^c\hat{\varphi}_{_L}
+\mu_{_L}\hat{\Phi}_{_L}\hat{\varphi}_{_L}\;,
\nonumber\\
&&{\cal W}_{_X}=\lambda_1\hat{Q}\hat{Q}_{_5}^c\hat{X}+\lambda_2\hat{U}^c\hat{U}_{_5}\hat{X}^\prime
+\lambda_3\hat{D}^c\hat{D}_{_5}\hat{X}^\prime+\mu_{_X}\hat{X}\hat{X}^\prime\;.
\label{superpotential}
\end{eqnarray}

The soft breaking terms are given by
\begin{eqnarray}
{\cal L}_{_{soft}}
&=&{\cal L}_{_{soft}}^{MSSM}-(m_{_{\tilde{N}^c}}^2)_{_{IJ}}\tilde{N}_I^{c*}\tilde{N}_J^c
-m_{_{\tilde{Q}_4}}^2\tilde{Q}_{_4}^\dagger\tilde{Q}_{_4}-m_{_{\tilde{U}_4}}^2\tilde{U}_{_4}^{c*}\tilde{U}_{_4}^c
-m_{_{\tilde{D}_4}}^2\tilde{D}_{_4}^{c*}\tilde{D}_{_4}^c
\nonumber\\
&&
-m_{_{\tilde{Q}_5}}^2\tilde{Q}_{_5}^{c\dagger}\tilde{Q}_{_5}^c-m_{_{\tilde{U}_5}}^2\tilde{U}_{_5}^*\tilde{U}_{_5}
-m_{_{\tilde{D}_5}}^2\tilde{D}_{_5}^*\tilde{D}_{_5}-m_{_{\tilde{L}_4}}^2\tilde{L}_{_4}^\dagger\tilde{L}_{_4}
-M_{_{\tilde{\nu}_4}}^2\tilde{\nu}_{_4}^{c*}\tilde{\nu}_{_4}^c-m_{_{\tilde{E}_4}}^2\tilde{e}_{_4}^{c*}\tilde{e}_{_4}^c
\nonumber\\
&&
-m_{_{\tilde{L}_5}}^2\tilde{L}_{_5}^{c\dagger}\tilde{L}_{_5}^c
-M_{_{\tilde{\nu}_5}}^2\tilde{\nu}_{_5}^*\tilde{\nu}_{_5}-m_{_{\tilde{E}_5}}^2\tilde{e}_{_5}^*\tilde{e}_{_5}
-m_{_{\Phi_{_B}}}^2\Phi_{_B}^*\Phi_{_B}-m_{_{\varphi_{_B}}}^2\varphi_{_B}^*\varphi_{_B}
\nonumber\\
&&
-m_{_{\Phi_{_L}}}^2\Phi_{_L}^*\Phi_{_L}
-m_{_{\varphi_{_L}}}^2\varphi_{_L}^*\varphi_{_L}-(m_{_B}\lambda_{_B}\lambda_{_B}
+m_{_L}\lambda_{_L}\lambda_{_L}+h.c.)
\nonumber\\
&&
+\Big\{A_{_{u_4}}Y_{_{u_4}}\tilde{Q}_{_4}H_{_u}\tilde{U}_{_4}^c+A_{_{d_4}}Y_{_{d_4}}\tilde{Q}_{_4}H_{_d}\tilde{D}_{_4}^c
+A_{_{u_5}}Y_{_{u_5}}\tilde{Q}_{_5}^cH_{_d}\tilde{U}_{_5}+A_{_{d_5}}Y_{_{d_5}}\tilde{Q}_{_5}^cH_{_u}\tilde{D}_{_5}
\nonumber\\
&&
+A_{_{BQ}}\lambda_{_Q}\tilde{Q}_{_4}\tilde{Q}_{_5}^c\Phi_{_B}+A_{_{BU}}\lambda_{_U}\tilde{U}_{_4}^c\tilde{U}_{_5}\varphi_{_B}
+A_{_{BD}}\lambda_{_D}\tilde{D}_{_4}^c\tilde{D}_{_5}\varphi_{_B}+B_{_B}\mu_{_B}\Phi_{_B}\varphi_{_B}
+h.c.\Big\}
\nonumber\\
&&
+\Big\{A_{_{e_4}}Y_{_{e_4}}\tilde{L}_{_4}H_{_d}\tilde{E}_{_4}^c+A_{_{N_4}}Y_{_{N_4}}\tilde{L}_{_4}H_{_u}\tilde{N}_{_4}^c
+A_{_{e_5}}Y_{_{e_5}}\tilde{L}_{_5}^cH_{_u}\tilde{E}_{_5}+A_{_{N_5}}Y_{_{\nu_5}}\tilde{L}_{_5}^cH_{_d}\tilde{N}_{_5}
\nonumber\\
&&
+A_{_N}Y_{_N}\tilde{L}H_{_u}\tilde{N}^c+A_{_{N^c}}\lambda_{_{N^c}}\tilde{N}^c\tilde{N}^c\varphi_{_L}
+B_{_L}\mu_{_L}\Phi_{_L}\varphi_{_L}+h.c.\Big\}
\nonumber\\
&&
+\Big\{A_1\lambda_1\tilde{Q}\tilde{Q}_{_5}^cX+A_2\lambda_2\tilde{U}^c\tilde{U}_{_5}X^\prime
+A_3\lambda_3\tilde{D}^c\tilde{D}_{_5}X^\prime+B_{_X}\mu_{_X}XX^\prime+h.c.\Big\}.
\end{eqnarray}
The first term ${\cal L}_{_{soft}}^{MSSM}$ denotes the soft breaking terms of MSSM.
To break the gauge symmetry from $SU(3)_C \otimes SU(2)_L \otimes U(1)_Y \otimes U(1)_B \otimes U(1)_L$
to electromagnetic symmetry $U(1)_e$,
nonzero VEVs $v_u, v_d$ and $v_B, \bar{v}_B, v_L,\bar{v}_L$
are allocated to
$SU(2)_L$ doublets $H_u, H_d$
and $SU(2)_L$ singlets $\Phi_B, \varphi_B, \Phi_L, \varphi_L$.

\begin{eqnarray}
H_u&=&\left(\begin{array}{c}
H_u^+\\ (v_u+H_u^0+iP_u^0)/\sqrt{2}
\end{array}\right),\nonumber\\
H_u&=&\left(\begin{array}{c}
(v_d+H_d^0+iP_d^0)/\sqrt{2}\\ H_d^-
\end{array}\right),\nonumber\\
\Phi_B &=&(v_B+\Phi_B^0+iP_B^0)/\sqrt{2},\nonumber\\
\varphi_B &=&(\bar{v}_B+\varphi_B^0+i\bar{P}_B^0)/\sqrt{2},\nonumber\\
\Phi_L &=&(v_L+\Phi_L^0+iP_L^0)/\sqrt{2},\nonumber\\
\varphi_L &=&(\bar{v}_L+\varphi_L^0+i\bar{P}_L^0)/\sqrt{2}.
\end{eqnarray}
Here we take the notation $\tan\beta=v_u/v_d, \tan\beta_B=\bar{v}_B/v_B$ and $\tan\beta_L=\bar{v}_L/v_L$.
After spontaneously breaking and unitary transformation from interactive eigenstate to mass eigenstate,
one can extract the Feynman rules and mass spectrums in BLMSSM.
The mass matrices of the particles
that mediate the one-loop process $b\to s\gamma$
can be found in ref. \cite{mass-matrice}.
The Feynman rules that we need can be extracted from the following terms,
where all the repeated index of generation should be summed over.
\begin{eqnarray}
&&\mathcal{L}_{H^\pm du}=\Big(-Y_d^IZ_H^{1i}P_L+Y_u^JZ_H^{2i}P_R\Big)K^{JI*}\bar{d}^Iu^JH_i^-,\nonumber\\
&&\mathcal{L}_{\tilde{D} \chi^0d}=\Big[\Big(\frac{-e}{\sqrt{2}s_Wc_W}Z_D^{Ii}
		(\frac{1}{3}Z_N^{1j}s_W-Z_N^{2j}c_W)+Y_d^IZ_D^{(I+3)i}Z_N^{3j}\Big)P_L\nonumber\\
          &&  \;\;\;\;\;\;\;\;\;\;\;\;\;\;\;
          +\Big(\frac{-e\sqrt{2}}{3c_W}Z_D^{(I+3)i}Z_N^{1j*}+Y_d^IZ_D^{Ii}Z_N^{3j*}
          \Big)P_R\Big] \bar{\chi}^0_jd^I\tilde{D}^+_i,\nonumber\\
&&\mathcal{L}_{\tilde{D}\chi_B^0d}=\frac{\sqrt{2}}{3}g_B\Big(Z_{N_B}^{1j}Z_D^{Ii}P_L
		+Z_{N_B}^{1j*}Z_D^{(I+3)i}P_R\Big)\bar{\chi}_{B_j}^0d^I\tilde{D}^+,\nonumber\\
&&\mathcal{L}_{\tilde{U}\chi^-d}=\Big[\big(\frac{-e}{s_W}Z_U^{Ji*}Z_+^{1j}
	+Y_u^JZ_U^{(J+3)i*}Z_+^{2j}\big)P_L-Y_d^IZ_U^{Ji*}Z_-^{2j*}P_R\Big]K^{JI}\bar{\chi}^-d\tilde{U}^-,\nonumber\\
&&\mathcal{L}_{Xb^\prime d}=\Big[\lambda_{1}(W_b^{\dag})_{j1}(Z_X)_{1k}P_L
	-\lambda_{3}^*(U_b^{\dag})_{j2}(Z_X)_{2k}P_R\Big]\bar{b}^\prime_jd^IX_k,\nonumber\\
&&\mathcal {L}_{\tilde{b}^\prime\tilde{X} d}
	=-\Big[\lambda_1(W_{\tilde{b}^\prime}^*)_{3\rho}P_L
	+\lambda_3^*(W_{\tilde{b}^\prime})_{4\rho}P_R\Big]
     \bar{\tilde{X}}d^I\tilde{b}^\prime_\rho,\nonumber\\
&&\mathcal {L}_{\tilde{D}\Lambda_G d}=g_3\sqrt{2}Y_{\alpha\beta}^a
	\big(-Z_D^{Ii}P_L+Z_D^{(I+3)i}P_R\big)\bar{\Lambda}_G^a d_\beta^I\tilde{D}^+_{i\alpha}.
\end{eqnarray}

\section{One-loop corrections to $b\rightarrow s\gamma$ from BLMSSM}
\label{one-loop-corrections}

The flavor transition process $b\rightarrow s\gamma$ can be described by effective Hamiltonian
at scale $\mu=O(m_b)$ as follow \cite{hamiltonian}:

\begin{eqnarray}
\mathcal {H}_{eff}(b\to s\gamma)=-\frac{4G_F}{\sqrt{2}}V_{ts}^*V_{tb}
\Big[C_1Q_1^c+C_2Q_2^c+\sum_{i=3}^6C_iQ_i+\sum_{i=7}^8(C_iQ_i+\tilde{C}_i\tilde{Q}_i)\Big],
\label{eff-hamiltonian}
\end{eqnarray}
and the operators are given by ref. \cite{operators1,operators2,operators3}:
\begin{eqnarray}
\mathcal{O}_1^c
	&=&(\bar{s}_L\gamma_\mu T^ab_L)(\bar{c}_L\gamma^\mu T^ab_L),\nonumber\\
\mathcal{O}_2^c
	&=&(\bar{s}_L\gamma_\mu b_L)(\bar{c}_L\gamma^\mu T^a b_L),\nonumber\\
\mathcal{O}_3
	&=&(\bar{s}_L\gamma_\mu b_L)\sum_q(\bar{q}\gamma^\mu q),\nonumber\\
\mathcal{O}_4
	&=&(\bar{s}_L\gamma_\mu T^ab_L)\sum_q(\bar{q}\gamma^\mu T^a q),\nonumber\\
\mathcal{O}_5
	&=&(\bar{s}_L\gamma_\mu \gamma_\nu\gamma_\rho b_L)\sum_q(\bar{q}\gamma^\mu \gamma^\mu \gamma^\nu\gamma^\rho q),\nonumber\\
\mathcal{O}_6
	&=&(\bar{s}_L\gamma_\mu \gamma_\nu\gamma_\rho T^a b_L)\sum_q(\bar{q}\gamma^\mu \gamma^\mu \gamma^\nu\gamma^\rho T^a q),\nonumber\\
\mathcal{O}_7
	&=&e/g_s^2m_b(\bar{s}_L\sigma_{\mu\nu}b_R)F^{\mu\nu},\nonumber\\
\mathcal{O}_8
	&=&1/g_s^2m_b(\bar{s}_L\sigma_{\mu\nu}T^ab_R)G^{a,\mu\nu},\nonumber\\
\tilde{\mathcal{O}}_7
	&=&e/g_s^2m_b(\bar{s}_R\sigma_{\mu\nu}b_L)F^{\mu\nu},\nonumber\\
\tilde{\mathcal{O}}_8
	&=&1/g_s^2m_b(\bar{s}_R\sigma_{\mu\nu}T^ab_L)G^{a,\mu\nu}.
\end{eqnarray}
Coefficients of these operators can be extracted from
Feynman amplitudes that originate from considered diagrams.
Actually only the Coefficients of $\mathcal{O}_{7,8}$ and $\tilde{\mathcal{O}}_{7,8}$
are needed if we adopt the branching ratio formula presented in ref. \cite{hamiltonian}:
\begin{eqnarray}
&& BR(\bar{B}\rightarrow X_s\gamma)_{NP}\nonumber\\
&=&10^{-4}\times\left\{(3.36\pm0.23)+\frac{16\pi^2a_{77}}{\alpha_s^2(\mu_b)}\big[|C_{7,NP}(\mu_{EW})|^2+|\tilde{C}_{7,NP}(\mu_{EW})|^2\big]\right.\nonumber\\
&&+\frac{16\pi^2a_{88}}{\alpha_s^2(\mu_b)}\big[|C_{8,NP}(\mu_{EW})|^2+|\tilde{C}_{8,NP}(\mu_{EW})|^2\big]\nonumber\\
&&+\frac{ 4\pi}{\alpha_s(\mu_b)}\mbox{Re}\big[a_7C_{7,NP}(\mu_{EW})+a_8C_{8,NP}(\mu_{EW})\big.\nonumber\\
&&+\left.\big.\frac{ 4\pi a_{78}}{\alpha_s(\mu_b)}\big(C_{7,NP}(\mu_{EW})C_{8,NP}(\mu_{EW})+\tilde{C}_{7,NP}(\mu_{EW})\tilde{C}_{8,NP}(\mu_{EW})\big)\big]\right\},
\end{eqnarray}
where the first term is SM prediction.
The others come from new physics in which
$C_{7,NP}(\mu_{EW})$, $C_{8,NP}(\mu_{EW})$, $\tilde{C}_{7,NP}(\mu_{EW})$ and $\tilde{C}_{8,NP}(\mu_{EW})$
indicate Wilson coefficients at electroweak scale.
It is an advantage of this expression that we don't have to evolve them down to hadronic scale $\mu\sim m_b$
as the effect of evolution has already been involved in the coefficients $a_{7,8,77,88,78}$.
The numerical values of these coefficients are given in table~\ref{coeff-a}.

\begin{table}[tbh]
\caption{Numerical values for the coefficients $a_{7,8,77,88,78}$ at electroweak scale.}
\begin{tabular}{@{}c|c|c|c|c@{}} \toprule
$a_7$ & $a_8$ & $a_{77}$ & $a_{88}$ & $a_{78}$\\
\colrule
$-7.184+0.612i$ & $-2.225-0.557i$ & $4.743$ & $0.789$ & $2.454-0.884i$\\
\botrule
\end{tabular}
\label{coeff-a}
\end{table}

\begin{figure}[tbh]
\centering 
\includegraphics[width=.8\textwidth,]{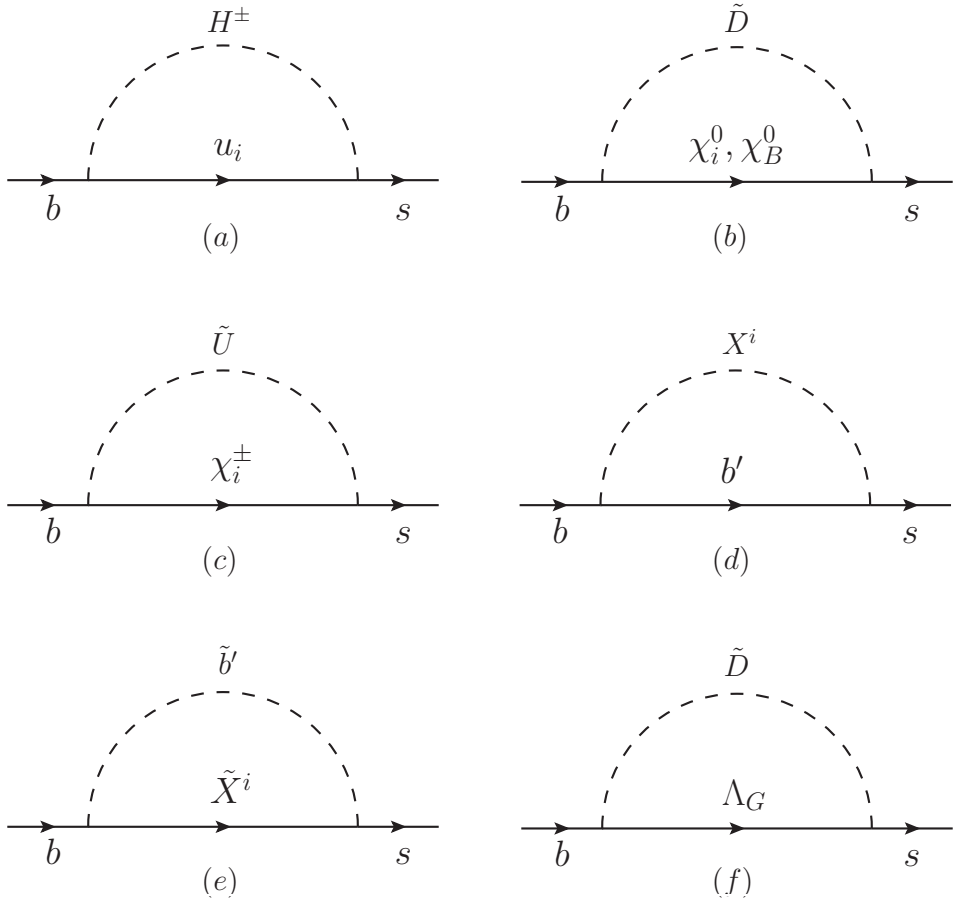}
\caption{ One-loop Feynman diagrams of $b\to s$. The inner-line particles $\chi_B^0, X$
denote baryon neutralinos and new scalar particle introduced in BLMSSM.
$b'$ and $\tilde{b}'$ are exotic quarks and squarks respectively.
The photon and gluon can be attached in all possible ways.}
\label{feyn-diag}
\end{figure}

To obtain the New Physics corrections in BLMSSM, we investigate one-loop diagrams shown in Figure~\ref{feyn-diag}.
Photons should be attached to all inner lines with electric charge to complete the diagrams of $b\to s\gamma$
that contribute to $\mathcal{O}_7$ and $\tilde{\mathcal{O}}_7$.
Similarly, diagrams of $b\to sg$ can be completed with gluons attached to all the inner lines with color charge,
and $\mathcal{O}_8$ and $\tilde{\mathcal{O}}_8$ originate from these process.

In details, we attach a photon to SM quark $u_i,(i=1,2,3)$
or charged Higgs $H^\pm$ in Figure~\ref{feyn-diag}.(a) to get a set of trigonal diagrams for $b\to s\gamma$,
while gluon can only be attached to up-type quarks $u_i$ to form a specific diagram of $b\to sg$.
To give a complete correction originating from Figure~\ref{feyn-diag}.(a),
contributions from all generations of $u_i$ and Higgs should be summed over.
From the amplitudes of these diagrams, one can extract Wilson coefficients of
electric- and chromomagnetic-dipole operators $\mathcal{O}_{7}$ and $\tilde{\mathcal{O}}_{7}$
at electroweak scale,
\begin{eqnarray}
 \frac{G_F}{\sqrt{2}}C_{7\gamma}^a(\Lambda)
&=&-i\Lambda^{-2}(V^*_{ts}V_{tb})^{-1}
	\left\{        (\eta^L_{H^\pm})_{su_i}^\dag(\eta^L_{H^\pm})_{u_ib}F_{1,\gamma}^{(a)}(x_{u_i},x_{H^\pm})\right.\nonumber\\
&&  \left.+\frac{m_f}{m_b}(\eta^L_{H^\pm})_{su_i}^\dag(\eta^R_{H^\pm})_{u_ib}F_{2,\gamma}^{(a)}(x_{u_i},x_{H^\pm})\right\},\nonumber\\
\frac{G_F}{\sqrt{2}}\tilde{C}_{7\gamma}^a(\Lambda)
&=& \frac{G_F}{\sqrt{2}}C_{7\gamma}^a(\Lambda)\big(\eta^L_{H^\pm}\leftrightarrow\eta^R_{H^\pm}\big),
\end{eqnarray}
where $x_i=m_i^2/\mu_{EW}^2$. The concrete expressions of relevant couplings are already given in previous section,
and the form factors can be written as:
\begin{eqnarray}
F_{1,\gamma}^{(a)}(x,y)&=&\Big[\frac{1}{72}\frac{\partial^3\varrho_{_{3,1}}}{\partial y^3}
                              +\frac{1}{24}\frac{\partial^2\varrho_{_{2,1}}}{\partial y^2}
                              -\frac{1}{ 6}\frac{\partial\varrho_{_{1,1}}}{\partial y}\Big](x,y),\nonumber\\
F_{2,\gamma}^{(a)}(x,y)&=&\Big[\frac{1}{12}\frac{\partial^2\varrho_{_{2,1}}}{\partial y^2}
                              -\frac{1}{6}\frac{\partial\varrho_{_{1,1}}}{\partial y}
                              -\frac{1}{3}\frac{\partial\varrho_{_{1,1}}}{\partial x}\Big](x,y),
\end{eqnarray}
where function $\varrho_{m,n}(x,y)$ is defined as:
\begin{eqnarray}
\varrho_{_{m,n}}(x,y)={x^m\ln^nx-y^m\ln^ny\over x-y}.
\label{varrho}
\end{eqnarray}

Corrections from all the other diagrams to $C_{7\gamma}$ and $\tilde{C}_{7\gamma}$ can be obtained similarly.
In Figure~\ref{feyn-diag}.(b), the photon can only be attached to charged -1/3 squark $\tilde{D}$.
We present contributions from both neutralinos $\chi_i^0$ and baryon neutralinos $\chi_B^0$ at electroweak scale as
\begin{eqnarray}
\frac{G_F}{\sqrt{2}}C_{7\gamma}^b(\Lambda)
   &=&-i\Lambda^{-2}(V^*_{ts}V_{tb})^{-1}
     \left\{        (\xi^L_{\chi_i^0})_{s\tilde{D}}^\dag(\xi^L_{\chi_i^0})_{\tilde{D}b}F_{1,\gamma}^{(b)}(x_{\chi_i^0},x_{\tilde{D}})
    +\frac{m_f}{m_b}(\xi^L_{\chi_i^0})_{s\tilde{D}}^\dag(\xi^R_{\chi_i^0})_{\tilde{D}b}F_{2,\gamma}^{(b)}(x_{\chi_i^0},x_{\tilde{D}})\right.\nonumber\\
   & &\;\;\;\; +\left.(\xi^L_{\chi_B^0})_{s\tilde{D}}^\dag(\xi^L_{\chi_B^0})_{\tilde{D}b}F_{1,\gamma}^{(b)}(x_{\chi_B^0},x_{\tilde{D}})
      +\frac{m_f}{m_b}(\xi^L_{\chi_B^0})_{s\tilde{D}}^\dag(\xi^R_{\chi_B^0})_{\tilde{D}b}F_{2,\gamma}^{(b)}(x_{\chi_B^0},x_{\tilde{D}})\right\},\nonumber\\
\frac{G_F}{\sqrt{2}}\tilde{C}_{7\gamma}^b(\Lambda)
   &=& \frac{G_F}{\sqrt{2}}C_{7\gamma}^b(\Lambda)\big(\xi^L_{\chi_i^0}\leftrightarrow\xi^R_{\chi_i^0}\;,\;\xi^L_{\chi_B^0}\leftrightarrow\xi^R_{\chi_B^0}\big).
\end{eqnarray}

\begin{eqnarray}
F_{1,\gamma}^{(b)}(x,y) &=& \big[-\frac{1}{72}\frac{\partial^3\varrho_{_{3,1}}}{\partial^3y}+\frac{1}{24}\frac{\partial^2\varrho_{_{2,1}}}{\partial^2y}\big](x,y),\nonumber\\
F_{2,\gamma}^{(b)}(x,y) &=& \big[-\frac{1}{12}\frac{\partial^2\varrho_{_{2,1}}}{\partial^2y}+\frac{1}{6}\frac{\partial  \varrho_{_{1,1}}}{\partial y}\big](x,y).
\end{eqnarray}

With the photon attached to the charged +2/3 squarks $\tilde{U}$ or chargino $\chi^\pm_i$ in Figure~\ref{feyn-diag}.(c), the contributions
to Wilson coefficients read
\begin{eqnarray}
\frac{G_F}{\sqrt{2}}C_{7\gamma}^c(\Lambda)
   &=&-i\Lambda^{-2}(V^*_{ts}V_{tb})^{-1}
     \left\{        (\eta^L_{\tilde{U}})_{s\chi_i^{\pm}}^\dag(\eta^L_{\tilde{U}})_{\chi_i^{\pm}b}F_{1,\gamma}^{(c)}(x_{\chi_i^{\pm}},x_{\tilde{U}})
    +\frac{m_f}{m_b}(\eta^L_{\tilde{U}})_{s\chi_i^{\pm}}^\dag(\eta^R_{\tilde{U}})_{\chi_i^{\pm}b}F_{2,\gamma}^{(c)}(x_{\chi_i^{\pm}},x_{\tilde{U}})\right\},\nonumber\\
\frac{G_F}{\sqrt{2}}\tilde{C}_{7\gamma}^c(\Lambda)
           &=& \frac{G_F}{\sqrt{2}}C_{7\gamma}^c(\Lambda)\big(\eta^L_{\tilde{U}}\leftrightarrow\eta^R_{\tilde{U}}\big).
\end{eqnarray}
\begin{eqnarray}
F_{1,\gamma}^{(c)}(x,y) &=& \big[-\frac{1}{72}\frac{\partial^3\varrho_{_{3,1}}}{\partial^3y}
                       +\frac{1}{ 6}\frac{\partial^2\varrho_{_{2,1}}}{\partial^2y}
                       -\frac{1}{ 4}\frac{\partial  \varrho_{_{1,1}}}{\partial  y}\big](x,y),\nonumber\\
F_{2,\gamma}^{(c)}(x,y) &=& \big[-\frac{1}{12}\frac{\partial^2\varrho_{_{2,1}}}{\partial^2y}
                                 +\frac{1}{ 6}\frac{\partial\varrho_{_{1,1}}}{\partial y}
                                 -\frac{1}{ 2}\frac{\partial\varrho_{_{1,1}}}{\partial x}\big](x,y).
\end{eqnarray}

The intermediate particles in Figure~\ref{feyn-diag}.(d) are the exotic quarks $b^\prime$ with charge -1/3 and superfield $X$ introduced in BLMSSM.
The contributions from this diagram are

\begin{eqnarray}
\frac{G_F}{\sqrt{2}}C_{7\gamma}^d(\Lambda)
   &=&-i\Lambda^{-2}(V^*_{ts}V_{tb})^{-1}
     \left\{        (\eta^L_{X^j})_{sb^\prime}^\dag(\eta^L_{X^j})_{b^\prime b}F_{1,\gamma}^{(d)}(x_{b^\prime},x_{X^j})
    +\frac{m_f}{m_b}(\eta^L_{X^j})_{sb^\prime}^\dag(\eta^R_{X^j})_{b^\prime b}F_{2,\gamma}^{(d)}(x_{b^\prime},x_{X^j})\right\},\nonumber\\
\frac{G_F}{\sqrt{2}}\tilde{C}_{7\gamma}^d(\Lambda)
           &=& \frac{G_F}{\sqrt{2}}C_{7\gamma}^d(\Lambda)\big(\eta^L_{X^j}\leftrightarrow\eta^R_{X^j}\big).
\end{eqnarray}

Correspondingly, the corrections of exotic squarks $\tilde{b}^\prime$ with charge -1/3 and fermionic particle $X$ can be obtained from Figure~\ref{feyn-diag}.(e)
\begin{eqnarray}
\frac{G_F}{\sqrt{2}}C_{7\gamma}^e(\Lambda)
   &=&-i\Lambda^{-2}(V^*_{ts}V_{tb})^{-1}
     \left\{ (\eta^L_{\tilde{b}^\prime})_{s\tilde{X}^j}^\dag(\eta^L_{\tilde{b}^\prime})_{\tilde{X}^jb}F_{1,\gamma}^{(e)}
     	(x_{\tilde{X}^j},x_{\tilde{b}^\prime})
    +\frac{m_f}{m_b}(\eta^L_{\tilde{b}^\prime})_{s\tilde{X}^j}^\dag(\eta^R_{\tilde{b}^\prime})_{\tilde{X}^jb}F_{2,\gamma}^{(e)}
    		(x_{\tilde{X}^j},x_{\tilde{b}^\prime})\right\},\nonumber\\
\frac{G_F}{\sqrt{2}}\tilde{C}_{7\gamma}^e(\Lambda)
           &=& \frac{G_F}{\sqrt{2}}C_{7\gamma}^e(\Lambda)\big(\eta^L_{\tilde{b}^\prime}\leftrightarrow\eta^R_{\tilde{b}^\prime}\big).
\end{eqnarray}

\begin{eqnarray}
F_{1,\gamma}^{(e)}(x,y) &=& \big[-\frac{1}{72}\frac{\partial^3\varrho_{_{3,1}}}{\partial^3y}+\frac{1}{24}\frac{\partial^2\varrho_{_{2,1}}}{\partial^2y}\big](x,y),\nonumber\\
F_{2,\gamma}^{(e)}(x,y) &=& \big[-\frac{1}{12}\frac{\partial^2\varrho_{_{2,1}}}{\partial^2y}+\frac{1}{ 6}\frac{\partial\varrho_{_{1,1}}}{\partial y}\big](x,y).
\end{eqnarray}

From Figure~\ref{feyn-diag}.(f), we obtain the corrections from gluinos $\Lambda_G$ in MSSM, the Wilson coefficients at $\mu_{EW}$ are
\begin{eqnarray}
\frac{G_F}{\sqrt{2}}C_{7\gamma}^f(\Lambda)
   &=&-i\Lambda^{-2}(V^*_{ts}V_{tb})^{-1}
     \left\{        (\eta^L_{\tilde{D}})_{s\Lambda_G}^\dag(\eta^L_{\tilde{D}})_{\Lambda_Gb}F_{1,\gamma}^{(f)}(x_{\Lambda_G},x_{\tilde{D}})
    +\frac{m_f}{m_b}(\eta^L_{\tilde{D}})_{s\Lambda_G}^\dag(\eta^R_{\tilde{D}})_{\Lambda_Gb}F_{2,\gamma}^{(f)}(x_{\Lambda_G},x_{\tilde{D}})\right\},\nonumber\\
\frac{G_F}{\sqrt{2}}\tilde{C}_{7\gamma}^f(\Lambda)
           &=& \frac{G_F}{\sqrt{2}}C_{7\gamma}^f(\Lambda)\big(\eta^L_{\tilde{D}}\leftrightarrow\eta^R_{\tilde{D}}\big).
\end{eqnarray}
with
\begin{eqnarray}
F_{1,\gamma}^{(f)}(x,y) &=& \big[\frac{1}{24}\frac{\partial^3\varrho_{_{3,1}}}{\partial^3y}-\frac{1}{8}\frac{\partial^2\varrho_{_{2,1}}}{\partial^2y}\big](x,y),\nonumber\\
F_{2,\gamma}^{(f)}(x,y) &=& \big[\frac{1}{ 4}\frac{\partial^2\varrho_{_{2,1}}}{\partial^2y}-\frac{1}{2}\frac{\partial\varrho_{_{1,1}}}{\partial y}\big](x,y).
\end{eqnarray}

The corrections to $C_{8g}$ and $\tilde{C}_{8g}$ at electroweak scale
can be obtained by attaching the gluon to intermediate virtual particles with colors.
For diagrams in Figure~\ref{feyn-diag}, the gluon can be attached to SM up-type quarks $u_i$, squarks in MSSM $\tilde{U}, \tilde{D}$,
exotic quarks $b^\prime$ with charge -1/3 and its supersymmetric partners $\tilde{b}^\prime$, as well as the gluinos $\Lambda_G$.
Wilson coefficients at electroweak scale can be formulated as:
\begin{eqnarray}
\frac{G_F}{\sqrt{2}}C_{8G}^a(\Lambda)    &=&-i\Lambda^{-2}(V^*_{ts}V_{tb})^{-1}
         \left\{    (\eta^L_{H^\pm})_{su_i}^\dag(\eta^L_{H^\pm})_{u_ib}F_{1,g}^{(a)}(x_{u_i},x_{H^\pm})
    +\frac{m_f}{m_b}(\eta^L_{H^\pm})_{su_i}^\dag(\eta^R_{H^\pm})_{u_ib}F_{2,g}^{(a)}(x_{u_i},x_{H^\pm})\right\},\nonumber\\
\frac{G_F}{\sqrt{2}}\tilde{C}_{8G}^a(\Lambda)&=& \frac{G_F}{\sqrt{2}}C_{8G}^a(\Lambda)
                                                (\eta^L_{H^\pm} \leftrightarrow \eta^R_{H^\pm}\;,\; \eta^L_{G^\pm} \leftrightarrow \eta^R_{G^\pm} ),\nonumber\\
\frac{G_F}{\sqrt{2}}C_{8G}^b(\Lambda)
   &=& -i\Lambda^{-2}(V^*_{ts}V_{tb})^{-1}
     \left\{        (\xi^L_{\chi_i^0})_{s\tilde{D}}^\dag(\xi^L_{\chi_i^0})_{\tilde{D}b}F_{1,g}^{(b)}(x_{\chi_i^0},x_{\tilde{D}})
    +\frac{m_f}{m_b}(\xi^L_{\chi_i^0})_{s\tilde{D}}^\dag(\xi^R_{\chi_i^0})_{\tilde{D}b}F_{2,g}^{(b)}(x_{\chi_i^0},x_{\tilde{D}})\right.\nonumber\\
   & &\;\;\;\; +\left.(\xi^L_{\chi_B^0})_{s\tilde{D}}^\dag(\xi^L_{\chi_B^0})_{\tilde{D}b}F_{1,g}^{(b)}(x_{\chi_B^0},x_{\tilde{D}})
      +\frac{m_f}{m_b}(\xi^L_{\chi_B^0})_{s\tilde{D}}^\dag(\xi^R_{\chi_B^0})_{\tilde{D}b}F_{2,g}^{(b)}(x_{\chi_B^0},x_{\tilde{D}})\right\},\nonumber\\
\frac{G_F}{\sqrt{2}}\tilde{C}_{8G}^b(\Lambda)
   &=& \frac{G_F}{\sqrt{2}}C_{8g}^b(\Lambda)\big(\xi^L_{\chi_i^0}\leftrightarrow\xi^R_{\chi_i^0}\;,\;\xi^L_{\chi_B^0}\leftrightarrow\xi^R_{\chi_B^0}\big),\nonumber\\
\frac{G_F}{\sqrt{2}}C_{8G}^c(\Lambda)
   &=& -i\Lambda^{-2}(V^*_{ts}V_{tb})^{-1}
     \left\{        (\xi^L_{\chi_i^\pm})_{s\tilde{D}}^\dag(\xi^L_{\chi_i^\pm})_{\tilde{D}b}F_{1,g}^{(c)}(x_{\chi_i^\pm},x_{\tilde{U}})
    +\frac{m_f}{m_b}(\xi^L_{\chi_i^\pm})_{s\tilde{D}}^\dag(\xi^R_{\chi_i^\pm})_{\tilde{D}b}F_{2,g}^{(c)}(x_{\chi_i^\pm},x_{\tilde{U}})\right\},\nonumber\\
\frac{G_F}{\sqrt{2}}\tilde{C}_{8G}^c(\Lambda)
   &=& \frac{G_F}{\sqrt{2}}C_{8g}^c(\Lambda)\big(\xi^L_{\chi_i^\pm}\leftrightarrow\xi^R_{\chi_i^\pm}\big),\nonumber\\
\frac{G_F}{\sqrt{2}}C_{8G}^d(\Lambda)
   &=&-i\Lambda^{-2}(V^*_{ts}V_{tb})^{-1}
     \left\{ (\eta^L_{X^j})_{sb^\prime}^\dag(\eta^L_{X^j})_{b^\prime b}
     		F_{1,g}^{(d)}(x_{b^\prime},x_{X^j})
    +\frac{m_f}{m_b}(\eta^L_{X^j})_{sb^\prime}^\dag(\eta^R_{X^j})_{b^\prime b}
    			F_{2,g}^{(d)}(x_{b^\prime},x_{X^j})\right\},\nonumber\\
\frac{G_F}{\sqrt{2}}\tilde{C}_{8G}^d(\Lambda)
           &=& \frac{G_F}{\sqrt{2}}C_{8G}^d(\Lambda)\big(\eta^L_{X^j}\leftrightarrow\eta^R_{X^j}\big),\nonumber\\
\frac{G_F}{\sqrt{2}}C_{8G}^e(\Lambda)
   &=& -i\Lambda^{-2}(V^*_{ts}V_{tb})^{-1}
     \left\{        (\xi^L_{\tilde{b}^\prime})_{s\tilde{X}^j}^\dag(\xi^L_{\tilde{b}^\prime})_{\tilde{X}^jb}F_{1,g}^{(e)}
     	(x_{\tilde{X}^j},x_{\tilde{b}^\prime})
    +\frac{m_f}{m_b}(\xi^L_{\tilde{b}^\prime})_{s\tilde{X}^j}^\dag(\xi^R_{\tilde{b}^\prime})_{\tilde{X}^jb}F_{2,g}^{(e)}
    		(x_{\tilde{X}^j},x_{\tilde{b}^\prime})\right\},\nonumber\\
\frac{G_F}{\sqrt{2}}\tilde{C}_{8G}^e(\Lambda)
   &=& \frac{G_F}{\sqrt{2}}C_{8G}^e(\Lambda)\big(\xi^L_{\tilde{b}^\prime}\leftrightarrow\xi^R_{\tilde{b}^\prime})\big),
\end{eqnarray}
with the form factors listed below. As gluon can only be attached to intermediate fermion
$u_i$ and $b^\prime$ in Figure~\ref{feyn-diag}.(a) and \ref{feyn-diag}.(d), so the form factors have the same expressions.
While in Figure~\ref{feyn-diag}.(b), \ref{feyn-diag}.(c) and Figure~\ref{feyn-diag}.(e),
the gluon can only be attached to scalar particles.
Then form factors associated to these diagrams are the same.
By summing over the contributions to Wilson coefficients when gluon attached to $\Lambda_G$ and $\tilde{D}$,
we get form factors of Figure~\ref{feyn-diag}.(f).
\begin{eqnarray}
F_{1,g}^{(a)}(x,y)&=&F_{1,g}^{(d)}(x,y)=\Big[
                     -\frac{1}{24}\frac{\partial^3\varrho_{_{3,1}}}{\partial^3y}
                     +\frac{1}{ 4}\frac{\partial^2\varrho_{_{2,1}}}{\partial^2y}
                     -\frac{1}{ 4}\frac{\partial\varrho_{_{1,1}}}{\partial y}\Big](x,y),\nonumber\\
F_{2,g}^{(a)}(x,y)&=&F_{2,g}^{(d)}(x,y)=\Big[
                     -\frac{1}{4}\frac{\partial^2\varrho_{_{2,1}}}{\partial^2y}
                     +\frac{1}{2}\frac{\partial  \varrho_{_{1,1}}}{\partial  y}
                     -\frac{1}{2}\frac{\partial  \varrho_{_{1,1}}}{\partial  x}\Big](x,y),\nonumber\\
F_{1,g}^{(b)}(x,y)&=& F_{1,g}^{(c)}(x,y)=F_{1,g}^{(e)}(x,y)=\Big[
                      \frac{1}{24}\frac{\partial^3\varrho_{_{3,1}}}{\partial^3y}
                         -\frac{1}{8}\frac{\partial^2\varrho_{_{2,1}}}{\partial^2y}\Big](x,y),\nonumber\\
F_{2,g}^{(b)}(x,y)&=&F_{2,g}^{(c)}(x,y)=F_{2,g}^{(e)}(x,y)=\Big[
                      \frac{1}{4}\frac{\partial^2\varrho_{_{2,1}}}{\partial^2y}
                         -\frac{1}{2}\frac{\partial  \varrho_{_{1,1}}}{\partial y}\Big](x,y),\nonumber\\
F_{1,g}^{(f)}(x,y)&=& \Big[\frac{1}{8}\frac{\partial^2\varrho_{_{2,1}}}{\partial^2y}
                     -\frac{1}{4}\frac{\partial  \varrho_{_{1,1}}}{\partial  y}\Big](x,y),\nonumber\\
F_{2,g}^{(f)}(x,y)&=&\Big[-\frac{1}{2}\frac{\partial  \varrho_{_{1,1}}}{\partial  x}\Big](x,y).
\end{eqnarray}

The Wilson coefficients obtained above can also be used to
direct CP-violation in $\bar{B}\rightarrow X_s\gamma$
and the time-dependent CP-asymmetry in $B\rightarrow K^*\gamma$.
The direct CP-violation $A^{CP}_{\bar{B}\rightarrow X_s\gamma}$
and CP-asymmetry $S_{K^*\gamma}$ are defined in hadronic scale \cite{ACP-def1,ACP-def2,ACP-def3,ACP-def4,ACP-def5}
\begin{eqnarray}
A^{CP}_{\bar{B}\rightarrow X_s\gamma}
&=&\left.\frac{\Gamma(\bar{B}\rightarrow X_s\gamma)-\Gamma(B\rightarrow X_{\bar{s}}\gamma)}
        {\Gamma(\bar{B}\rightarrow X_s\gamma)+\Gamma(B\rightarrow X_{\bar{s}}\gamma)}\right|_{E_\gamma > (1-\delta)E_\gamma^{max}}\nonumber\\
&\simeq &\frac{10^{-2}}{|C_7(\mu_b)|^2}
    \Big[1.23\mathfrak{I}\big(C_2(\mu_b)C_7^*(\mu_b)\big)\nonumber\\
&&    -9.52\mathfrak{I}\big(C_8(\mu_b)C_7^*(\mu_b)\big)+0.01\mathfrak{I}\big(C_2(\mu_b)C_8^*(\mu_b)\big)\Big],\\
S_{K^*\gamma}
&\simeq &\frac{2\text{Im}(e^{-i\phi_d}C_7(\mu_b)C_7'(\mu_b))}{|C_7(\mu_b)|^2+|C_7'(\mu_b)|^2},
\end{eqnarray}
where the photon energy cut in $A^{CP}$ is taken as $\delta=3$,
and $\phi_d$ in $S_{K^*\gamma}$ is phase of $B_d$ mixing amplititude.
Here we use the experimental data $\sin\phi_d=0.67\pm0.02$ given in ref. \cite{1010-1589}.

As the Wilson coefficients are calculated at electroweak scale $\mu_{EW}$,
we need to evolve them down to hadronic scale $\mu\sim m_b$ with renormalization group equations.
\begin{eqnarray}
\vec{C}_{NP}(\mu)&=&\hat{U}(\mu,\mu_0)\vec{C}_{NP}(\mu_0),\nonumber\\
\vec{C}'_{NP}(\mu)&=&\hat{U}'(\mu,\mu_0)\vec{C}'_{NP}(\mu_0),
\label{evolve-equ}
\end{eqnarray}
where the Wilson coefficients are constructed as
\begin{eqnarray}
\vec{C}^\mathrm{T}_{NP}&=&(C_{1,NP},\cdots,C_{6,NP},C^{eff}_{7,NP},C^{eff}_{8,NP}),\nonumber\\
\vec{C}^{\prime,\mathrm{T}}_{NP}&=&(C'^{eff}_{7,NP},C'^{eff}_{8,NP}).
\end{eqnarray}
To be convenient, we take the numerrical results of effective coefficients $C^{eff}$ from SM at next-to-next-to-leading logarithmic(NNLL)
level, $C_7^{eff}(m_b)=-0.304, C_8^{eff}(m_b)=-0.167$.
The evolving matrices involved in Eq.(\ref{evolve-equ}) are given as
\begin{eqnarray}
\hat{U}(\mu,\mu_0)\simeq 1- \left[ \frac{1}{2\beta_0}\ln\frac{\alpha_s(\mu)}{\alpha_s(\mu_0)} \right]\hat{\gamma}^{(0)T},\nonumber\\
\hat{U}'(\mu,\mu_0)\simeq 1- \left[ \frac{1}{2\beta_0}\ln\frac{\alpha_s(\mu)}{\alpha_s(\mu_0)} \right]\hat{\gamma}'^{(0)T},
\end{eqnarray}
with anomalous dimension matrices
\begin{eqnarray}
\hat{\gamma}^{(0)}
=\left(\begin{array}{cccccccc}
-4 & \frac{8}{3} & 0 & -\frac{2}{9} & 0 & 0 & -\frac{208}{243} & \frac{173}{162}\\
12 & 0 & 0 & \frac{4}{3} & 0 & 0 & \frac{416}{81} & \frac{70}{27}\\
0 & 0 & 0 &  -\frac{52}{3} & 0 & 2 & -\frac{176}{81} & \frac{14}{27}\\
0 & 0 & -\frac{40}{9} & -\frac{100}{9} & \frac{4}{9} & \frac{5}{6} & -\frac{152}{243} & -\frac{587}{162}\\
0 & 0 & 0 & -\frac{256}{3} & 0 & 20 & -\frac{6272}{81} & \frac{6596}{27}\\
0 & 0 & -\frac{256}{9} & \frac{56}{9} & \frac{40}{9} & -\frac{2}{3} & \frac{4624}{243} & \frac{4772}{81}\\
0 & 0 & 0 & 0 & 0 & 0 & \frac{32}{3} & 0\\
0 & 0 & 0 & 0 & 0 & 0 & -\frac{32}{9} & \frac{28}{3}
\end{array}\right),
\end{eqnarray}
and
\begin{eqnarray}
\hat{\gamma}'^{(0)}
=\left(\begin{array}{cc}
\frac{32}{3} & 0\\
-\frac{32}{9} & \frac{28}{3}
\end{array}\right).
\end{eqnarray}

\section{Numerical analysis}
\label{sec-numericial}
The consistency of SM prediction and experimental data on $\bar{B}\to X_s\gamma$
sets stringent constraint on new physics parameters.
In this section, we discuss the numerical results of branching ratio with some assumptions.
The SM inputs are given in Table \ref{SM-inputs}.
All the parameters with mass dimension are given in the unit GeV.
To be concise, we omit all the unit GeV in this section.
Other free parameters introduced in BLMSSM
are set to be
$A_{BU}=A_{BD}=A_{BQ}=A'_{BU}=A'_{BD}=A'_{BQ}
=A_{d_4}=A_{d_5}=A_{u_4}=A_{u_5}
=A'_{d_4}=A'_{d_5}=A'_{u_4}=A'_{u_5}=100$,
$M^2_{\tilde{Q}_4}=M^2_{\tilde{Q}_5}
=M^2_{\tilde{U}_4}=M^2_{\tilde{U}_5}
=M^2_{\tilde{D}_4}=M^2_{\tilde{D}_5}=2500$,
$m_1=m_2=1200$ and $m_{Z_B}=1000$
to make sure the masses of new physics particles under experimental limitations.

\begin{figure}[htbp]
\centering
\subfigure[]{
\begin{minipage}[t]{0.5\linewidth}
\centering
\includegraphics[width=3.5in]{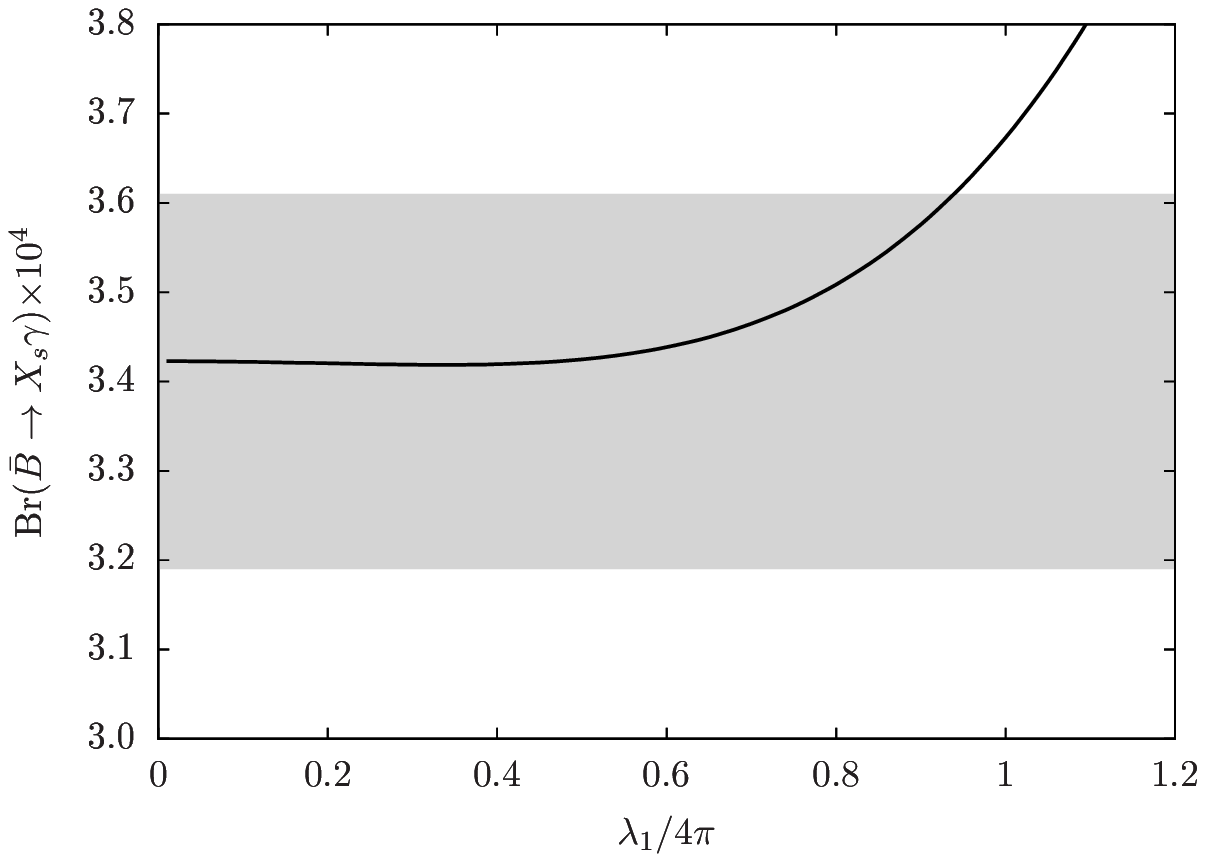}
\end{minipage}%
}%
\subfigure[]{
\begin{minipage}[t]{0.5\linewidth}
\centering
\includegraphics[width=3.5in]{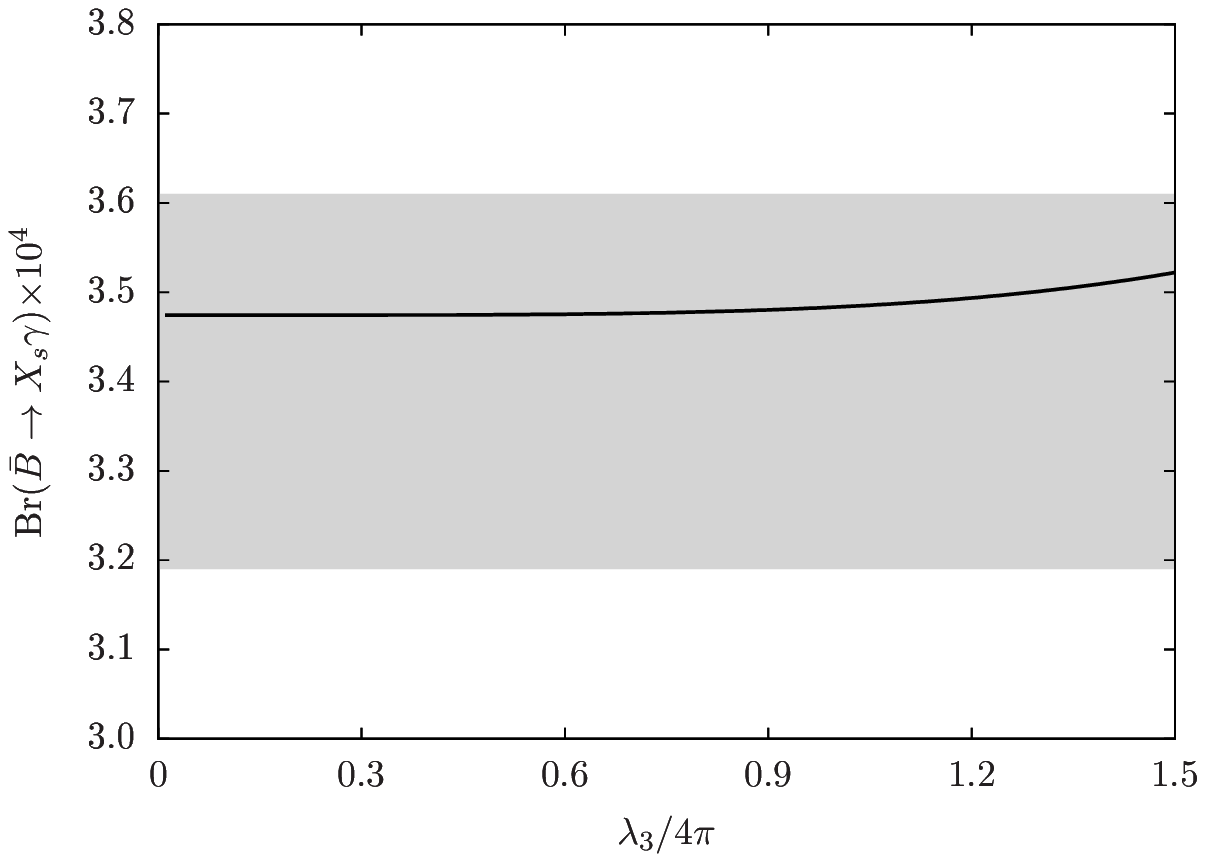}
\end{minipage}%
}%
\quad
\subfigure[]{
\begin{minipage}[t]{0.5\linewidth}
\centering
\includegraphics[width=3.5in]{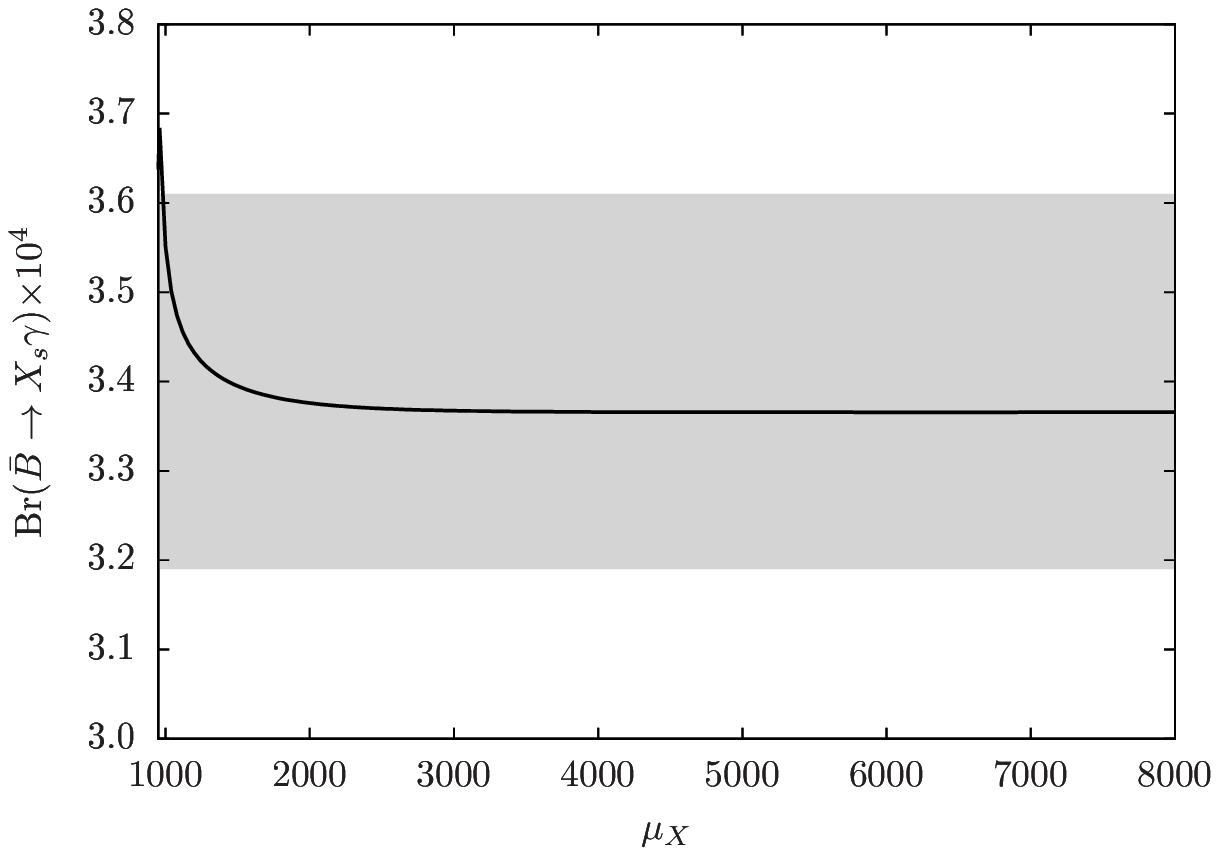}
\end{minipage}
}%
\subfigure[]{
\begin{minipage}[t]{0.5\linewidth}
\centering
\includegraphics[width=3.5in]{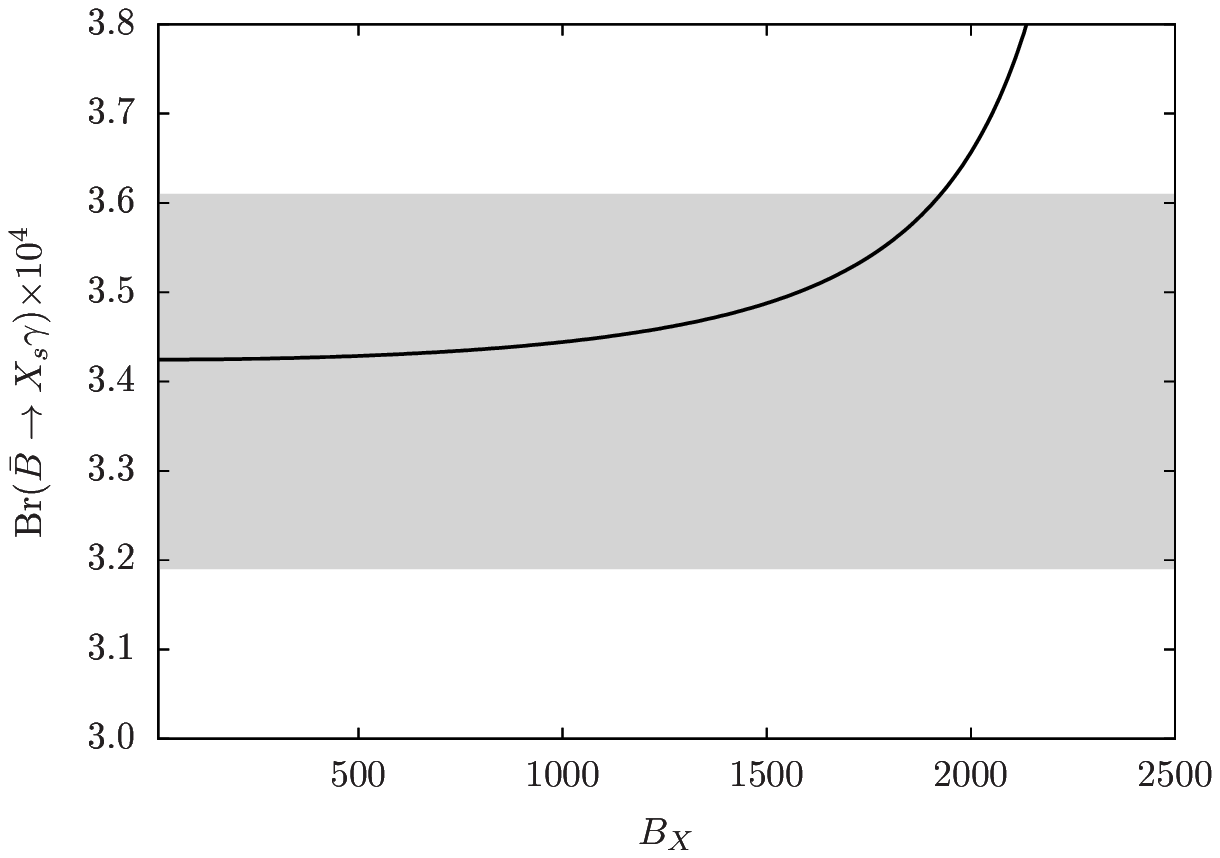}
\end{minipage}
}%
\centering
\caption{$Br(\bar{B}\to X_s\gamma)$ varying with parameters relevant to superfield $X$.}
\label{paras-X}
\end{figure}

As a new field introduced in BLMSSM, superfield $X$ interacts with exotic quarks.
The coplings between $X$ and $\hat{Q_5}, \hat{U_5}$
denoted by $\lambda_i, (i=1,2,3)$ are given in Eq.\ref{superpotential}.
From the analytical expressions, one can find the Wilson
coefficients are sensitive to these couplings as well as coefficients of mass term of $X$,
which turns up in $\mathcal{W}_X$ as $\mu_X$ and $B_X$.
We show the branching ratio varying with
$\lambda_1$, $\lambda_3$, $\mu_X$ and $B_X$ in firgure \ref{paras-X}.
The dependency of $\lambda_2$ is not listed as it is similar to $\lambda_1$.

\begin{table}[h]
\caption{\label{SM-inputs} SM inputs in numerical analysis.}
\centering
\begin{tabular}{|cc|cc|cc|}
\toprule
$\alpha$  & $1/128$ & $m_W$ & $80.385$ & $m_Z$ & $91.188$  \\ \colrule
$m_u$  & $0.0023$ & $m_c$ & $1.275$ & $m_t$ & $173.5$  \\   \colrule
$m_d$  & $0.0048$ & $m_s$ & $0.095$ & $m_b$ & $4.18$  \\
\botrule
\end{tabular}
\end{table}

In Figure \ref{paras-X}.(a), one can find the branching ratio increases when
$\lambda_1$ raises up.
The experimental limitations are denoted by the gray area,
and we have taken
$\lambda_3/(4\pi)=0.07, \tan\beta=10, \lambda_Q=0.7, \lambda_U=0.3, \lambda_D=0.2, \mu=-600, \mu_B=\mu_X=m_{Z_B}=m_B=1000, B_X=400, v_{bt}=6000$.
It can be seen in this figure that branching ratio reachs the upper limitations of experimental limitations when
$\lambda_1/(4\pi)\sim 0.94$,
then we get the constrain $\lambda_1/(4\pi)<0.94$.

Similarly, we plot the branching ratio varying with $\lambda_3$ in Figure \ref{paras-X}.(b).
By taking $\lambda_1/(4\pi)=0.07$, which satisfies the limitations obtained in Figure \ref{paras-X}.(a),
and $\tan\beta=5, \lambda_Q=0.2, \lambda_U=0.5, \lambda_D=0.8, \mu=-800, \mu_B=\mu_X=1100, m_{Z_B}=1000, m_B=500, B_X=400, v_{bt}=6000$,
we find the branching ratio rises very slowly when $\lambda_3/(4\pi)$ runs from 0.01 to 1.5.
The whole curve lies in the gray area, which means the branching ratio
satisfies the experimental constraint under our assumptions.

To investigate the trends of $Br(\Bar{B}\to X_s\gamma)$ varying with $\mu_X$,
we take $\lambda_1/(4\pi)=0.06, \lambda_3/(4\pi)=0.08,
\tan\beta=5, \lambda_Q=0.8, \lambda_U=0.5, \lambda_D=0.2,
\mu=-600, \mu_B=1000, B_X=400, m_{z_B}=1100, v_{bt}=6000, m_B=2500$.
We find from Figure \ref{paras-X}.(c) that the branching ratio diminishes steeply with increasing of $\mu_X$,
and finally gets to the value of standard model.
In Figure \ref{paras-X}.(d), we plot the branching ratio varying with $B_X$,
where we take $\lambda_1/(4\pi)=0.15, \lambda_3/(4\pi)=0.08,
\tan\beta=5, \lambda_Q=0.7, \lambda_U=0.2, \lambda_D=0.3,
\mu=-1000, \mu_B=1100, \mu_X=2500, m_{z_B}=900, v_{bt}=5500, m_B=2000$.
With the upper limitations of experimental result, we get the constraints
$B_X<1930$.

\begin{figure}[tbp]
\centering
\subfigure[]{
\begin{minipage}[t]{0.5\linewidth}
\centering
\includegraphics[width=3.2in]{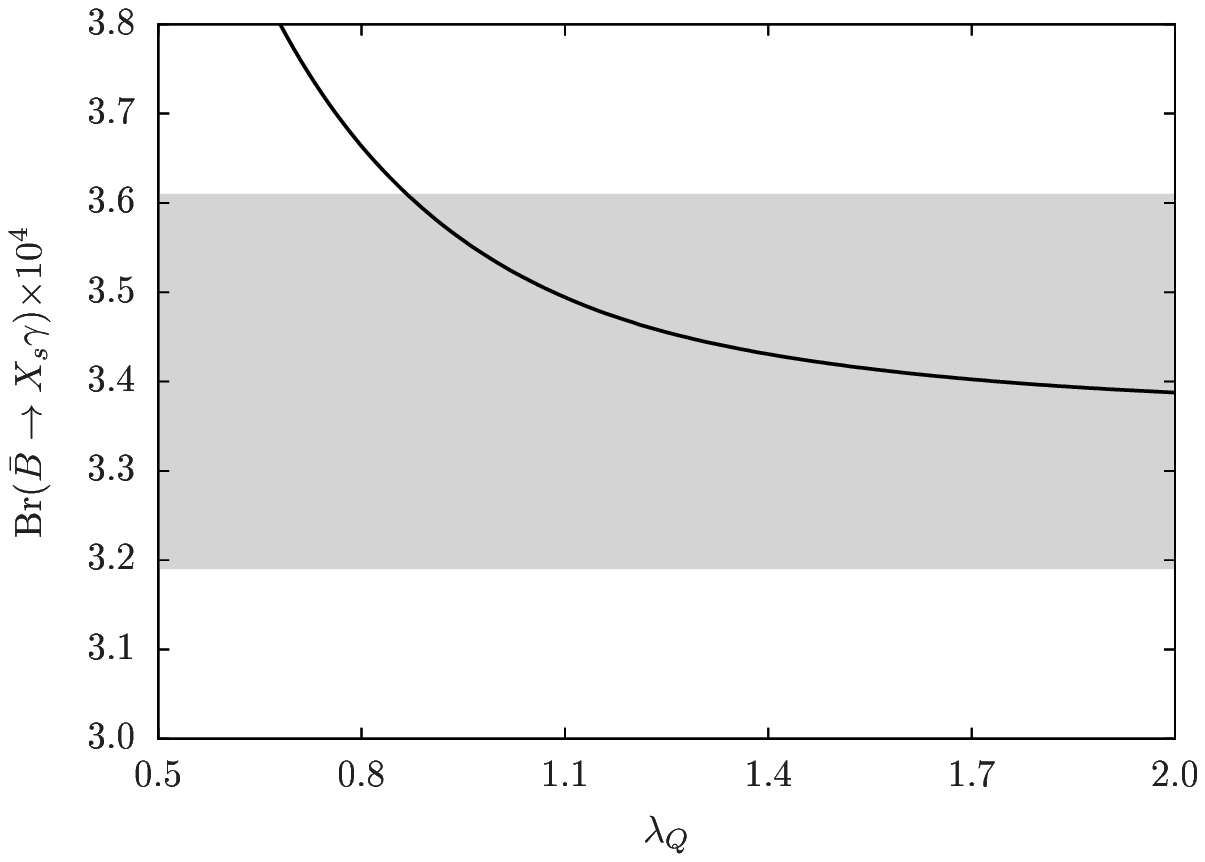}
\end{minipage}%
}%
\subfigure[]{
\begin{minipage}[t]{0.5\linewidth}
\centering
\includegraphics[width=3.2in]{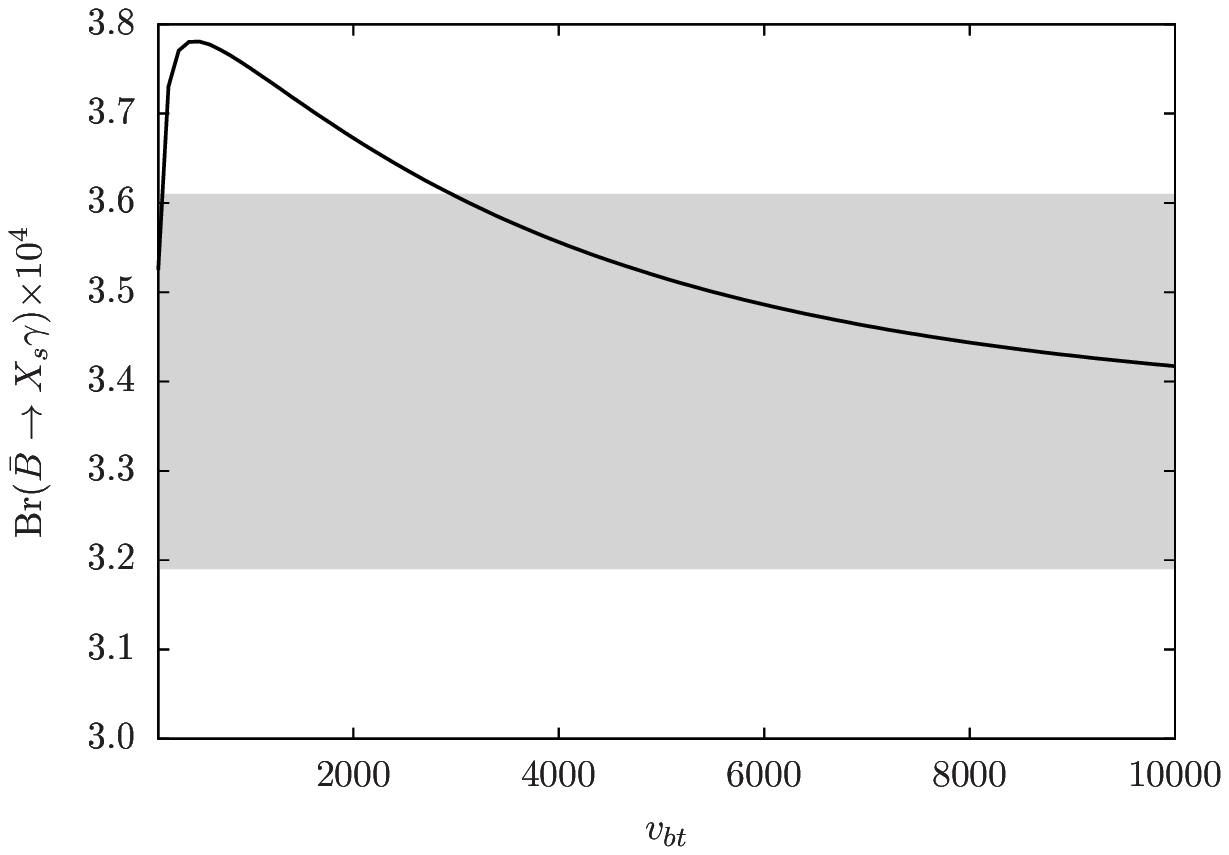}
\end{minipage}%
}%
\centering
\caption{$Br(\bar{B}\to X_s\gamma)$ varying with $\lambda_Q$ and $v_{bt}$.}
\label{br-para}
\end{figure}

In Figure \ref{br-para}.(a), we present the branching ratio varying with $\lambda_Q$,
which is the coupling truns up in superpotential term $\lambda_{_Q}\hat{Q}_{_4}\hat{Q}_{_5}^c\hat{\Phi}_{_B}$.
With $\lambda_1/(4\pi)=0.08, \lambda_3/(4\pi)=0.06,
\tan\beta=20, \lambda_U=0.3, \lambda_D=0.6,
\mu=-800, \mu_B=\mu_X=1000, B_X=400, m_{z_B}=1000, v_{bt}=5000, m_B=1500$,
we find branching ratio decreases when $\lambda_Q$ gets larger.
To consist with the experimental data,
one has $\lambda_Q>0.87$.
Another interesting parameter is $v_{bt}$, which is defined as
$v_{bt}=\sqrt{\bar{v}_B^2+v_B^2}$, where $v_B$ and $\bar{v}_B$
are VEVs of $\Phi_B$ and $\varphi_B$ respectively.
We plot branching ratio varying with $v_{bt}$ in Figure \ref{br-para}.(b) with
$\lambda_1/(4\pi)=\lambda_3/(4\pi)=0.1,
\tan\beta=5, \lambda_Q=0.4, \lambda_U=0.2, \lambda_D=0.7,
\mu=-1000, \mu_B=1100, \mu_X=1500, B_X=400, m_{z_B}=1000, m_B=1500$.
To satisfy the experimental constraints, we have $v_{bt}>2900$.

Additionally, we plot the direct CP-violation of $\bar{B}\to X_s\gamma$ and
time-dependent CP-asymmetry of $B\rightarrow K^*\gamma$ varying with
$\lambda_1, \lambda_3, \mu_X, B_X, \lambda_Q, \lambda_D$ and $v_{bt}$.
Within the framework of SM, we have
$-0.6\%<A_{CP}^{SM}<+2.8\%$ \cite{PRL-106-2011-141801},
and the average value of this observable is $A_{CP}^{exp}=-0.009\pm 0.018$\cite{PDG}.
Within some uncertainty, the theoretical value is consistent with the experimental result.
Compared with direct CP-violation of $\bar{B}\to X_s\gamma$,
there is significant deviation between
SM prediction and experimental result of $S_{K^*\gamma}$.
The SM prediction of time-dependent CP-asymmetry in $B\rightarrow K^*\gamma$ at LO level is
given as $S_{K^*\gamma}^{SM}\simeq (-2.3\pm1.6)\% $ \cite{skr-SM}
and the experimental result is $S_{K^*\gamma}\simeq -0.15\pm 0.22 $ \cite{PDG, operators1}.

\begin{figure}[tb]
\centering
\subfigure[]{
\begin{minipage}[t]{0.5\linewidth}
\centering
\includegraphics[width=3.5in]{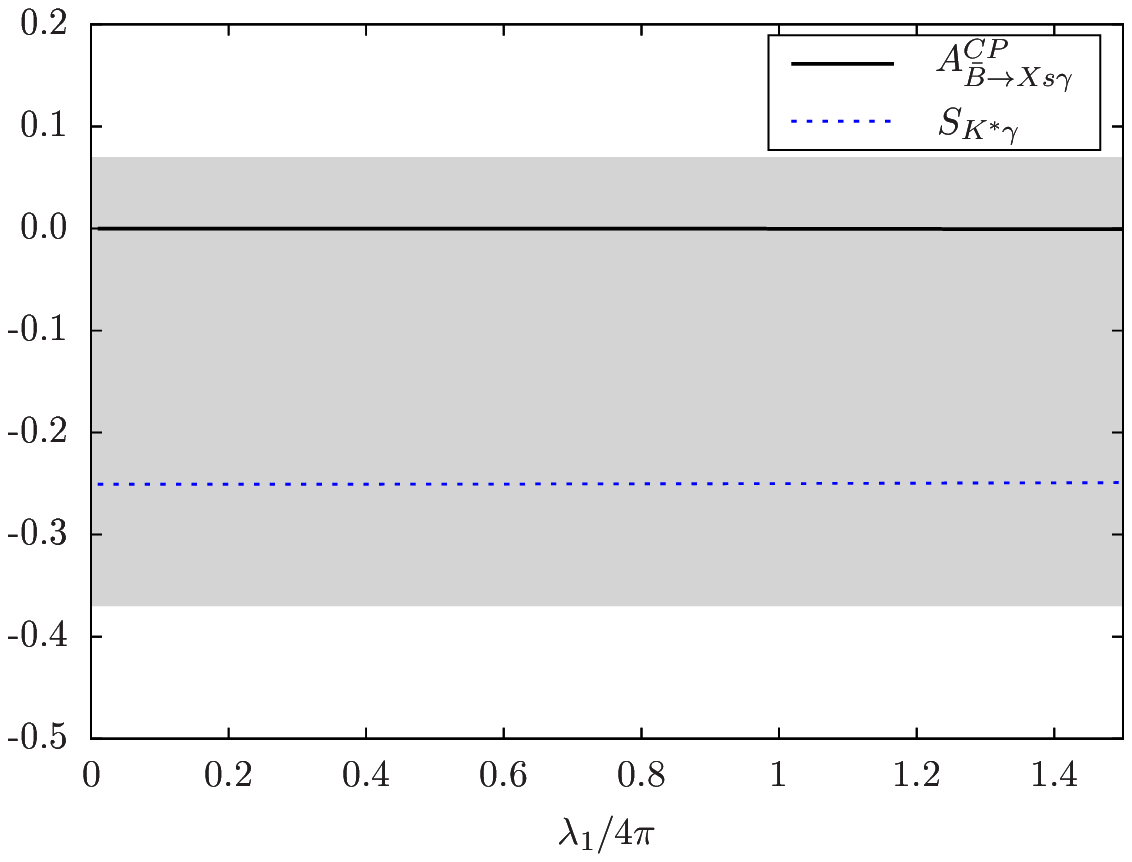}
\end{minipage}%
}%
\subfigure[]{
\begin{minipage}[t]{0.5\linewidth}
\centering
\includegraphics[width=3.5in]{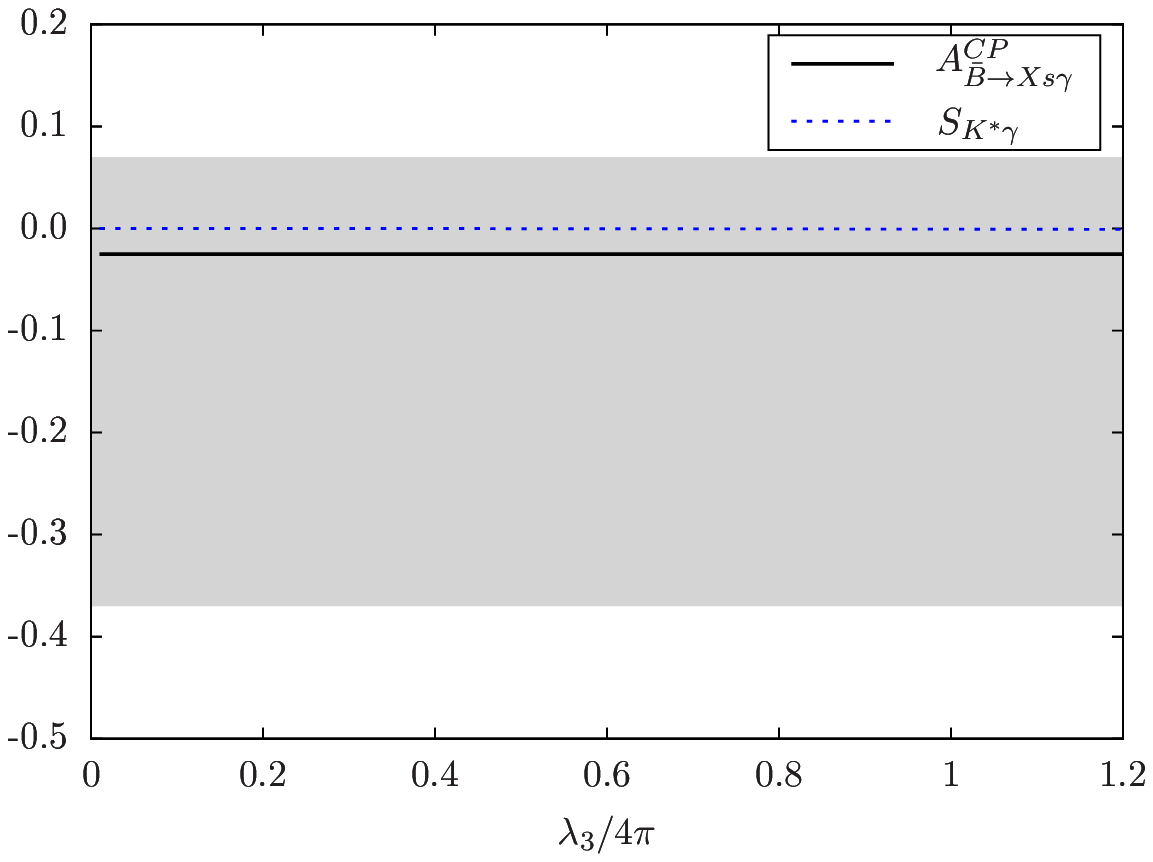}
\end{minipage}%
}%
\quad
\subfigure[]{
\begin{minipage}[t]{0.5\linewidth}
\centering
\includegraphics[width=3.5in]{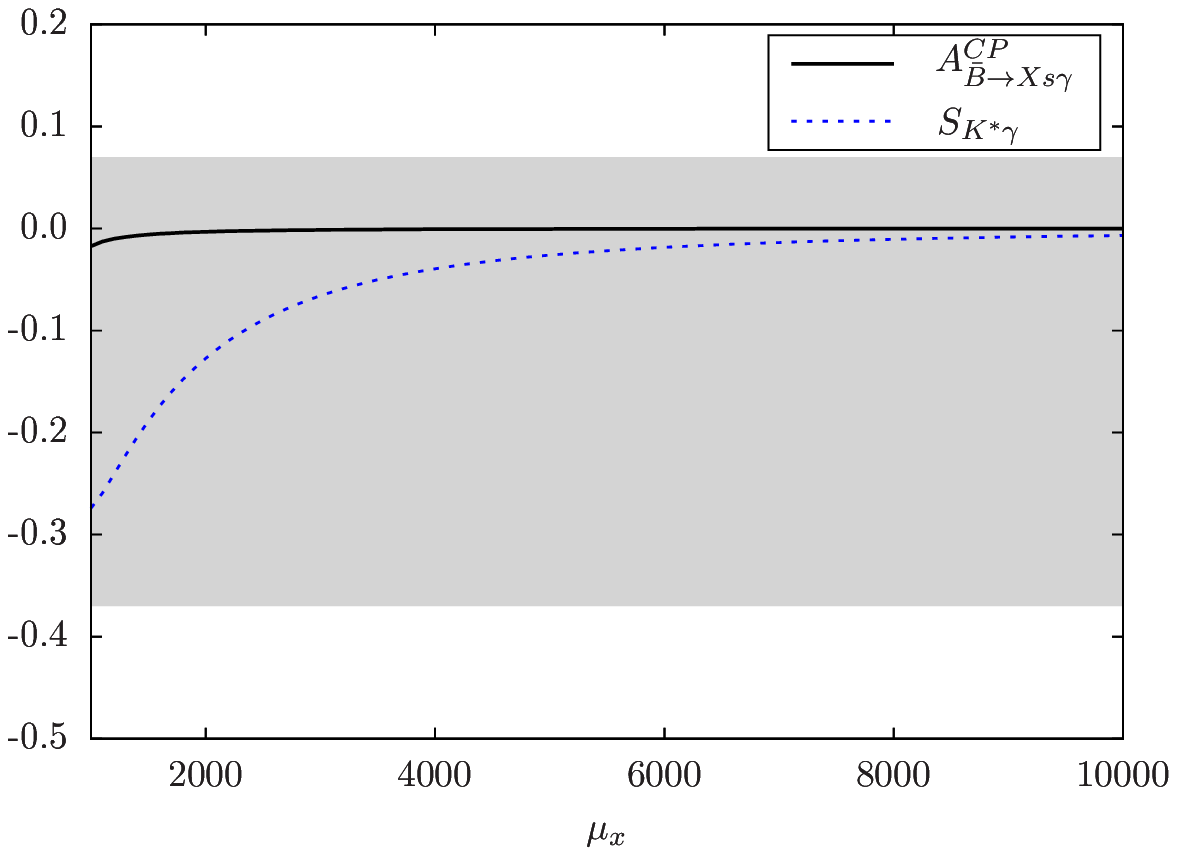}
\end{minipage}
}%
\subfigure[]{
\begin{minipage}[t]{0.5\linewidth}
\centering
\includegraphics[width=3.5in]{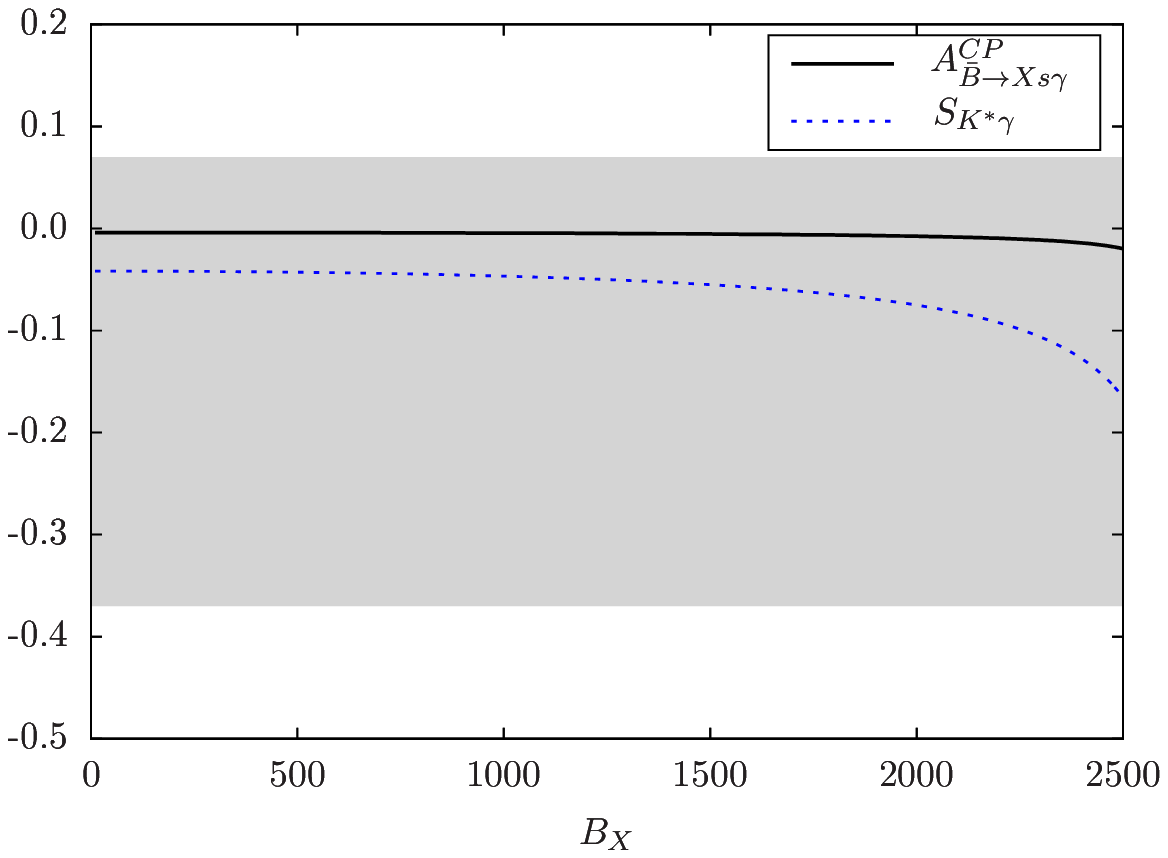}
\end{minipage}
}%
\centering
\caption{$A^{CP}_{\bar{B}\rightarrow X_s\gamma}$ and $S_{K^*\gamma}$ varying with paremeters relevant to superfield $X$.}
\label{CP-X}
\end{figure}

To investigate $A^{CP}_{\bar{B}\to X_s\gamma}$ and $S_{K^*\gamma}$ numerically,
some parameters are taken to be complex, and the area within experimental boundaries are filled to be gray
in the presented figures.
In Figure \ref{CP-X}, we plot the dependency of parameters relevant to superfield $X$.
Under our assumptions of free parameters introduced in BLMSSM,
we find that $A^{CP}_{\bar{B}\to X_s\gamma}$ (solid line) are hardly affected by the change of $\lambda_1, \lambda_3, \mu_X, B_X$.
Though corrections from one-loop level are almost zero,
the numerical results are consistent with experimental data.

As shown in Figure \ref{CP-X}.(a), one-loop corrections to $S_{K^*\gamma}$ (dashed line)
in BLMSSM can reach $-0.25$ with appropriate inputs.
By changing the free parameters, one finds $S_{K^*\gamma}$ can be as small as zero in Figure \ref{CP-X}.(b).
In Figure \ref{CP-X}.(c), it can be seen that $S_{K^*\gamma}$ raise obviously with increasing of $\mu_X$,
and finally gets stable around zero.
The $S_{K^*\gamma}$ varying with $B_X$ are given in Figure \ref{CP-X}.(d).
When $B_X$ raises up, we can see that $S_{K^*\gamma}$ decreases.
Within the range of parameters $\lambda_1, \lambda_3, \mu_X$ and $B_X$,
we find $S_{K^*\gamma}$ is consistent with experimental data.

\begin{figure}[tbp]
\centering
\subfigure[]{
\begin{minipage}[t]{0.5\linewidth}
\centering
\includegraphics[width=3.5in]{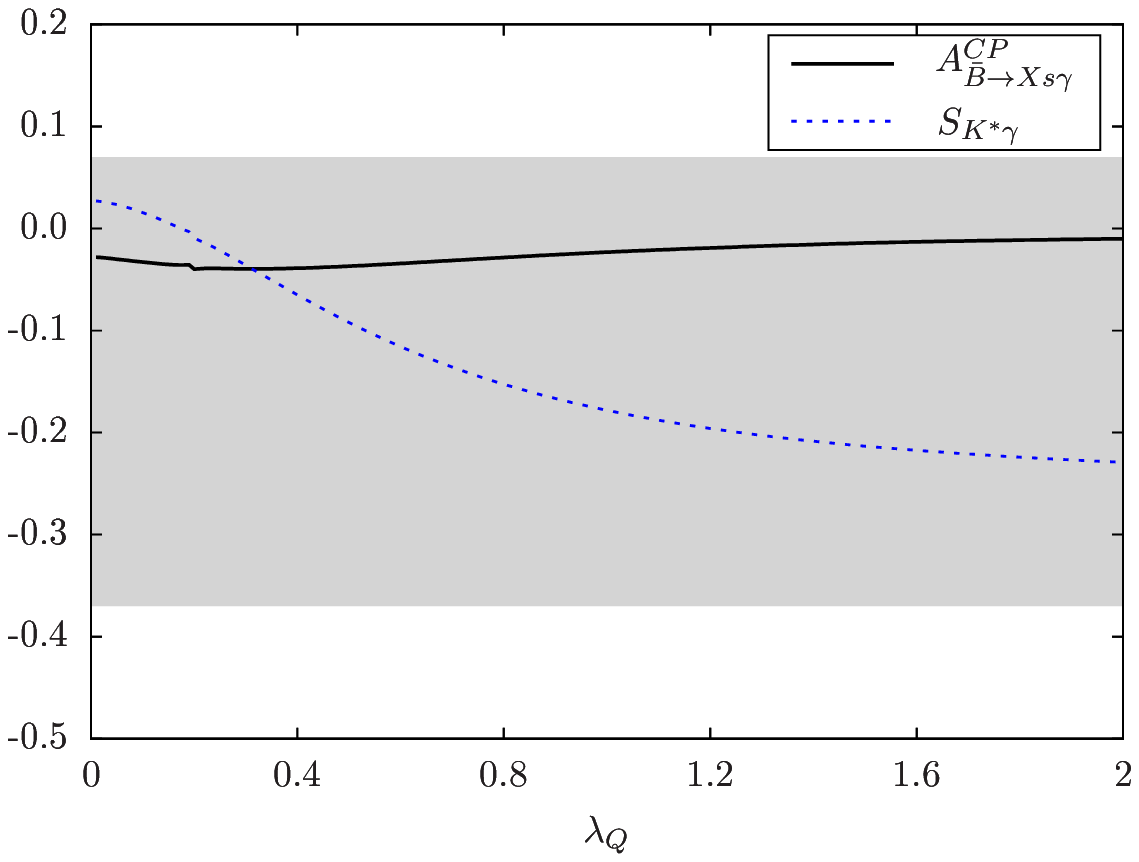}
\end{minipage}%
}%
\subfigure[]{
\begin{minipage}[t]{0.5\linewidth}
\centering
\includegraphics[width=3.5in]{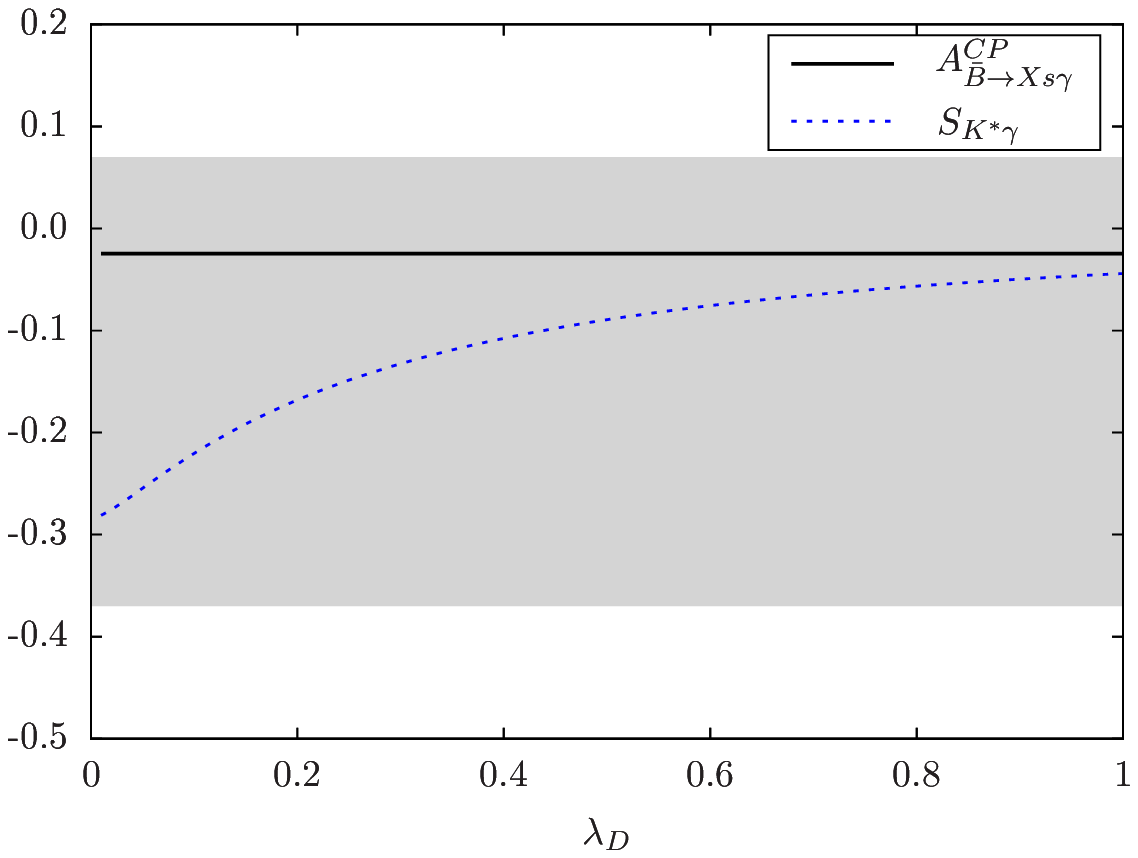}
\end{minipage}%
}%
\centering
\caption{$A^{CP}_{\bar{B}\rightarrow X_s\gamma}$ and $S_{K^*\gamma}$ varying with $\lambda_Q$ and $\lambda_D$.}
\label{CP-para}
\end{figure}

In Figure \ref{CP-para}, we take into account the parameters $\lambda_Q$ and $\lambda_D$.
When $\lambda_Q$ runs from 0.01 to 2.0, the time-dependent CP-asymmetry decrease from $0.02$ to $-0.22$.
While for the increasing of $\lambda_D$, $S_{K^*\gamma}$ raises from $-0.28$ to $-0.02$.
Under our assumptions, we conclude that $\lambda_Q$ and $\lambda_D$ affect
$S_{K^*\gamma}$ apparently, and the numerical results of new physics correction
are consistent with experimental data.
However, the direct CP-violation of $\bar{B}\to X_s\gamma$ depends on $\lambda_Q$ and $\lambda_D$ weakly,
and the one-loop contributions from BLMSSM are very small.

\begin{figure}[tb]
\centering 
\includegraphics[width=.6\textwidth]{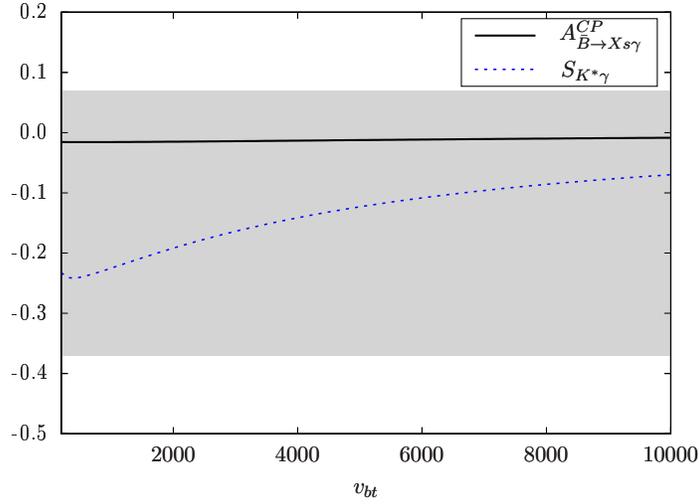}
\caption{$A^{CP}_{\bar{B}\rightarrow X_s\gamma}$ and $S_{K^*\gamma}$ varying with $v_{bt}$.}
\label{CP-Vbt}
\end{figure}

The last Figure \ref{CP-Vbt} illustrates the trend of $S_{K^*\gamma}$ and $A^{CP}_{\bar{B}\to X_s\gamma}$
varying with $v_{bt}$. By taking $\lambda_1/(4\pi)=0.8, \lambda_3/(4\pi)=0.9, B_X=400$ and $\lambda_Q=0.4e^{0.625\pi}$,
we find that $S_{K^*\gamma}$ increases from $-0.26$ to $-0.06$.
The $A^{CP}_{\bar{B}\to X_s\gamma}$ stays around zero within the range $100<v_{bt}<10000$.

\section{Conclusions}
\label{sec-Conclusions}
As an interesting process of FCNC, we investigate the transition $b\to s\gamma$ within the framework of BLMSSM.
With effective Hamiltonian method, we present the Wilson coefficients extracted from amplitudes
corresponding to the concerned one-loop diagrams.
Based on the analytical expressions, constraints on parameters are given in the numerical section
with the experimental data of branching ratio of $\bar{B}\to X_s\gamma$.
The direct CP-violation of $\bar{B}\to X_s\gamma$ in BLMSSM is very small,
and depend on the free parameters weakly.
However, the time-dependent CP-asymmetry $S_{K^*\gamma}$ in $B\rightarrow K^*\gamma$ varies with
$\mu_X, B_X, \lambda_Q, \lambda_D$ and $v_{bt}$ obviously.
The contributions from new physics can reach $-0.28$ under appropriate setup of the parameters.


\begin{thebibliography}{99}

\bibitem {PDG}
M. Tanabashi et al.(Particle Data Group),
Phys. Rev. D 98: 030001 (2018)

\bibitem{BrSM1}
M. Misiak et al.,
Phys. Rev. Lett. 114: 221801 (2015)

\bibitem{BrSM2}
M. Czakon, P. Fiedler, T. Huber, M. Misiak, T. Schutzmeier and M. Steinhauser,
(2015), arXiv:1503.01791[hep-ph]


\bibitem{BrSM3}
M. Misiak et al.,
Phys. Rev. Lett. 98: 022002 (2007)

\bibitem{BrSM4}
M. Misiak and M. Steinhauser,
Nucl. Phys. B 764: 62 (2007)

\bibitem{BrSM5}
K. Chetyrkin, M. Misiak and M. Munz,
Phys. Lett. B 400: 206 (1997)

\bibitem{BrSM6}
C. Greub, T. Hurth and D. Wyler,
Phys. Rev. D 54: 3350 (1996)

\bibitem{BrSM7}
K. Adel and Y.-P. Yao,
Phys. Rev. D 49: 4945 (1994)

\bibitem{BrSM8}
A. Ali and C. Greub,
Phys. Lett. B 361: 146 (1995)


\bibitem{MSSM}
J. Rosiek,
Phys. Rev. D 41: 3464 (1990)

\bibitem{BLMSSM1}
P. F. Perez and M. B. Wise,
JHEP, 1108: 068 (2011)

\bibitem{BLMSSM2}
P. F. Perez and M. B. Wise,
Phys. Rev. D 82: 011901 (2010)


\bibitem{PR597-2015-1-30}
P. F. Perez,
Physics reports 597: 1 (2015)


\bibitem{JHEP1411-119}
S. M. Zhao, T.F. Feng, H.B.Zhang, B. Yan and Y. J. Zhang,
JHEP,1411: 119 (2014)

\bibitem{EPJC-zhao-2017}
S. M. Zhao, T. F. Feng, Z, J. Yang, H. B. Zhang, X. X. Dong and T. Guo,
Eur. Phys. J. C77: 102 (2017)

\bibitem{NPB-871-2013-223}
T. F. Feng, S. M. Zhao, H. B. Zhang, Y. J. Zhang, Y. L. Yan,
Nucl. Phys. B 871: 223 (2013)


\bibitem{mass-matrice}
H. Li, J. B. Chen and L. L. Xing,
MPLA 33: 1850034 (2018)



\bibitem{hamiltonian}
E. Lunghi and J. Matias,
JHEP 0704:058 (2007)


\bibitem{operators1}
T. F. Feng, Y. L. Yan, H. B. Zhang and S. M. Zhao,
Phys. Rev. D 92 : 055024 (2015)

\bibitem{operators2}
C. Bobeth, M. Misiak and J. Urban,
Nucl. Phys. B 574:291-330 (2000)

\bibitem{operators3}
W. Altmannshofer, P. Ball1, A. Bharucha, A. J. Buras, D. M. Straub1 and M. Wick,
JHEP 01:019 (2009)


\bibitem{ACP-def1}
H. M. Asatrian and A. N. Ioannissian,
Phys. Rev. D 54 (1996) 5642

\bibitem{ACP-def2}
M. Ciuchini, E. Gabrielli and G.F. Giudice,
Phys. Lett. B388:353 (1996)

\bibitem{ACP-def3}
S. Baek, P. Ko,
Phys. Rev. Lett. 83:488 (1999)

\bibitem{ACP-def4}
A. L. Kagan and M. Neubert,
Phys. Rev. D 58:094012 (1998)

\bibitem{ACP-def5}
K. Kiers, A. Soni, and G. H. Wu,
Phys. Rev. D62:116004 (2000)


\bibitem{1010-1589}
D. Asner et al.(Heavy Flavor Averaging Group), (2010)
arXiv:1010.1589[hep-ex]

\bibitem{PRL-106-2011-141801}
M. Benzke, S. J. Lee, M. Neubert and G. Paz,
Phys. Rev. Lett. 106:141801 (2011)


\bibitem{skr-SM}
P. Ball, G. W. Jones and R. Zwicky
Phys. Rev. D 75:054004 (2007)

\end{thebibliography}
\end{document}